\documentclass[a4paper,11pt]{article}
\pdfoutput=1 

\usepackage{jcappub} 
\usepackage[T1]{fontenc} 
\usepackage{lineno}
\title{\boldmath Long-baseline horizontal radio-frequency transmission through polar ice} 

\author[2,3]{Allison, P.}
\author[4]{Archambault, S.}
\author[2,3]{Beatty, J.J.}
\author[1,7]{Besson, D.Z.}
\author[8]{Chen, C.C.}
\author[8]{Chen, C.H.}
\author[8]{Chen, P.}
\author[5]{Christenson, A.}
\author[2,3]{Clark, B.A.}
\author[10]{Clay, W.}
\author[2,3]{Connolly, A.}
\author[9]{Cremonesi, L.}
\author[10]{Deaconu, C.}
\author[5]{Duvernois, M.}
\author[6]{Friedman, L.}
\author[4]{Gaior, R.}
\author[2,3]{Hanson, J.}
\author[5]{Hanson, K.}
\author[5]{Haugen, J.}
\author[6]{Hoffman, K.D.}
\author[2,3]{Hong, E.}
\author[1]{Hornhuber, C.}
\author[8]{Hsu, S.Y.}
\author[8]{Hu, L.}
\author[8]{Huang, J.J.}
\author[8]{Huang, M.-H. A.}
\author[10]{Hughes, K.}
\author[4]{Ishihara, A.}
\author[5]{Karle, A.}
\author[5]{Kelley, J.L.}
\author[5]{Khandelwal, R.}
\author[4]{Kim, M.-C.}
\author[11]{Kravchenko, I.}
\author[11]{Kruse, J.}
\author[4]{Kurusu, K.}
\author[4]{Kuwabara, T.}
\author[1]{Latif, U.A.}
\author[5]{Laundrie, A.}
\author[8]{Li, C.-J.}
\author[8]{Liu, T.-C.}
\author[5]{Lu, M.-Y.}
\author[1]{Madison, B.}
\author[4]{Mase, K.}
\author[5]{Meures, T.}
\author[4]{Nam, J.}
\author[9]{Nichols, R.J.}
\author[1,7]{Novikov, A.}
\author[1]{Nozdrina, A.}
\author[10]{Oberla, E.}
\author[13]{Pan, Y.}
\author[2,3]{Pfendner, C.}
\author[3]{Relich, M.}
\author[5]{Sandstrom, P.}
\author[13]{Seckel, D.}
\author[8]{Shiao, Y.S.}
\author[1]{Shultz, A.}
\author[10]{Smith, D.}
\author[6]{Song, M.}
\author[2,3]{Torres, J.}
\author[6]{Touart, J.}
\author[14]{Varner, G.S.}
\author[10]{Vieregg, A.}
\author[8]{Wang, M.Z.}
\author[8]{Wang, S.H.}
\author[15]{Wissel, S.}
\author[4]{Yoshida, S.}
\author[1]{Young, R.}
\author[12]{Jordan, T.}
\affiliation[1]{Dept. of Physics and Astronomy, Univ. of Kansas, Lawrence, KS, USA}
\affiliation[2]{Dept. of Physics, The Ohio State University, 191 West Woodruff Avenue, Columbus, OH, USA}
\affiliation[3]{Center for Cosmology and Astro-Particle Physics, The Ohio State University, 191 West Woodruff Avenue, Columbus, OH, USA}
\affiliation[4]{Dept. of Physics, Chiba University, Tokyo, Japan}
\affiliation[5]{Dept. of Physics and Wisconsin IceCube Particle Astrophysics Center, University of Wisconsin, Madison, WI, USA}
\affiliation[6]{Dept. of Physics, Univ. of Maryland, College Park, MD, USA}
\affiliation[7]{National Research Nuclear University, Moscow Engineering Physics Institute, Moscow, Russia}
\affiliation[8]{Dept. of Physics, Grad. Inst. of Astrophys., \& Leung Center for Cosmology and Particle Astrophysics, National Taiwan Univ., Taipei, Taiwan}
\affiliation[9]{Dept. of Physics and Astronomy, Univ. College London, London, United Kingdom}
\affiliation[10]{Dept. of Physics, University of Chicago, Chicago, IL, USA}
\affiliation[11]{Dept. of Physics and Astronomy, Univ. of Nebraska-Lincoln, NE, USA}
\affiliation[13]{Dept. of Physics and Astronomy, Univ. of Delaware, Newark, DE, USA}
\affiliation[14]{Dept. of Physics and Astronomy, Univ. of Hawaii, Manoa, HI, USA}
\affiliation[15]{Dept. of Physics, California Polytechnic State University, San Luis Obispo, CA, USA}
\emailAdd{dbesson@ku.edu; dvbesson@mephi.ru}

\abstract{ 
We report on analysis of englacial radio-frequency (RF) pulser data received over horizontal baselines of 1--5 km, 
based on broadcasts from two sets of transmitters deployed to depths of up to 1500 meters at the South Pole. 
First, we analyze data collected using
two RF bicone transmitters 1400 meters below the ice surface, and
frozen into boreholes drilled for the IceCube experiment in 2011. 
Additionally, in Dec., 2018, a fat-dipole antenna, fed by one of three high-voltage (${\cal O}$(1 kV)), fast (${\cal O}$(1-5 ns risetime))
signal generators
was lowered into the 1700-m deep icehole drilled for the South Pole Ice Core Experiment (SPICE), approximately 3 km from the geographic South Pole. 
Signals from transmitters were recorded on the five englacial multi-receiver ARA stations, with receiver depths between 60--200 m. From analysis of deep transmitter
data, we estimate: i) the range of refractive index profiles of Antarctic ice with depth allowed by our measurements,  
ii) due to birefringence, a time difference between arrival times for 
vertically polarized vs. horizontally polarized signals (per km) for horizontally propagating signal, and
iii) for the first time, the attenuation length
for electromagnetic signals in the radio-frequency regime broadcast horizontally 
(rather than reflected vertically from bedrock). 
We additionally present data suggesting anomalous ice propagation effects, and contrary to expectations for
a transport medium with a smoothly varying refractive index profile.

Our results imply negligible uncertainty in overall neutrino detection volume due to refractive index uncertainties.
Our birefringence time-difference measurements are fit to the functional form
$\delta_t(H-V) [{\rm ns/km}]=a\cos\theta+b$, with $H/V$ the signal arrival times for the horizontally/vertically polarized EM signal components, and $\theta$ the 
opening angle in the horizontal plane between the signal propagation direction and the local ice flow direction, extracting
a=8.3$\pm$1.3 ns/km, and b=-8.6$\pm$0.9 ns/km (errors combined statistical and systematic), allowing a $\sim$15\% range estimate for future
measurements of in-ice neutrino interactions.
Finally, we find attenuation length values clustering around 
1.5 km, with measurements from the bicone transmitters yielding $L_{atten}=1.43\pm0.25\pm0.37$ km.
Taken together, these measurements support cold polar ice as a near-optimal platform for ultra-high energy neutrino detectors.
}

\begin{document}
\maketitle
\flushbottom

\section{Introduction}
Although originally purposed for radio-wave detection of ultra-high energy neutrinos interacting in cold polar ice via the Askaryan effect\cite{Askaryan1962a,Askaryan1962b,Askaryan1965}, the RICE\cite{KravchenkoFrichterSeckel2003}, ANITA\cite{GorhamAllisonBarwick2009}, ARA\cite{Allison:2011wk} and ARIANNA\cite{Barwick:2014rca} experiments have also realized extensive radioglaciological programs\cite{kravchenko2004situ,gorham2017hical,besson2008situ,besson2015antarctic,barwick2015radio,allison2019measurement}. Precise, and complete characterization of ice RF dielectric properties is important to optimizing both the layout as well as the trigger for existing, as well as planned neutrino detection experiments. Previous studies have measured the imaginary and real portions of the complex dielectric permittivity\cite{kravchenko2004situ,Besson:2010ww,barwick2005south}, based primarily on vertical radar echoes measured through South Polar ice, consistent with the typical, and logistically simplest, bistatic (i.e., nearly co-located receiver and transmitter) geometry. RF transmissions along a primarily horizontal chord require in-ice transmitter (Tx) and receiver (Rx) km-scale horizontal separations, and present a considerably greater logistical challenge. 
We herein detail our observations of horizontally-propagating radio-frequency signals from transmitters at depths up to 1400 meters in South Polar ice, and quantify, where possible, the associated permittivity. 

\subsection{Ice-based neutrino detectors}
The ARA, ARIANNA and ANITA experiments were designed to discover neutrinos at energies comparable to the ultra high energy charged cosmic rays (UHECR; $E>$10 PeV) detected by surface detector arrays such as the Telescope Array\cite{Abbasi:2014fya} and the Auger Experiment\cite{AugerNIM2004}. Such neutrinos carry unique information from corners of the Universe otherwise inaccessible to charged UHECR astroparticle physics, owing to the GZK-effect\cite{GZK1,GZK2,GZK3}. The experimental technique is based on measurement of radio-frequency coherent Cherenkov radiation emitted by the particle shower produced by an in-ice neutrino-molecular interaction\cite{Askaryan1962a,Askaryan1962b,Askaryan1965}. As part of the initial ARA receiver calibration plan one decade ago, three `deep' radio-frequency pulsers were deployed during the last year (2011-12) of ice-hole drilling for the IceCube experiment at the 1--2 km depths typical of neutrino interactions. Two of these transmitters were deployed at z=--1400 meter depths in IceCube holes 1 (transmitter `IC1S') and 22 (transmitter `IC22S'), respectively. A third was deployed below the IceCube array in hole 1 (`IC1D') to a depth of 2400 meters, but failed shortly after initial operation. Deployment of six ARA receiver stations (``testbed'' [2011], ``A1'' [2012], ``A2''/``A3'' [2013] and ``A4''/``A5'' [2017]), displaced 1-5 km horizontally from the IceCube array, allow study of radio signals propagating at primarily horizontal trajectories through the ice. Co-located with ARA station A5, the addition of a phased-array in the 2017-18 austral field season, with significantly lower signal thresholds\cite{allison2019design} constitutes a powerful augmentation to both the physics, as well as glaciological reach of the ARA instrument.

\subsection{Complex Dielectric Permittivity Measurements}
\begin{figure}[htpb]\includegraphics[angle=0,width=0.9\textwidth]{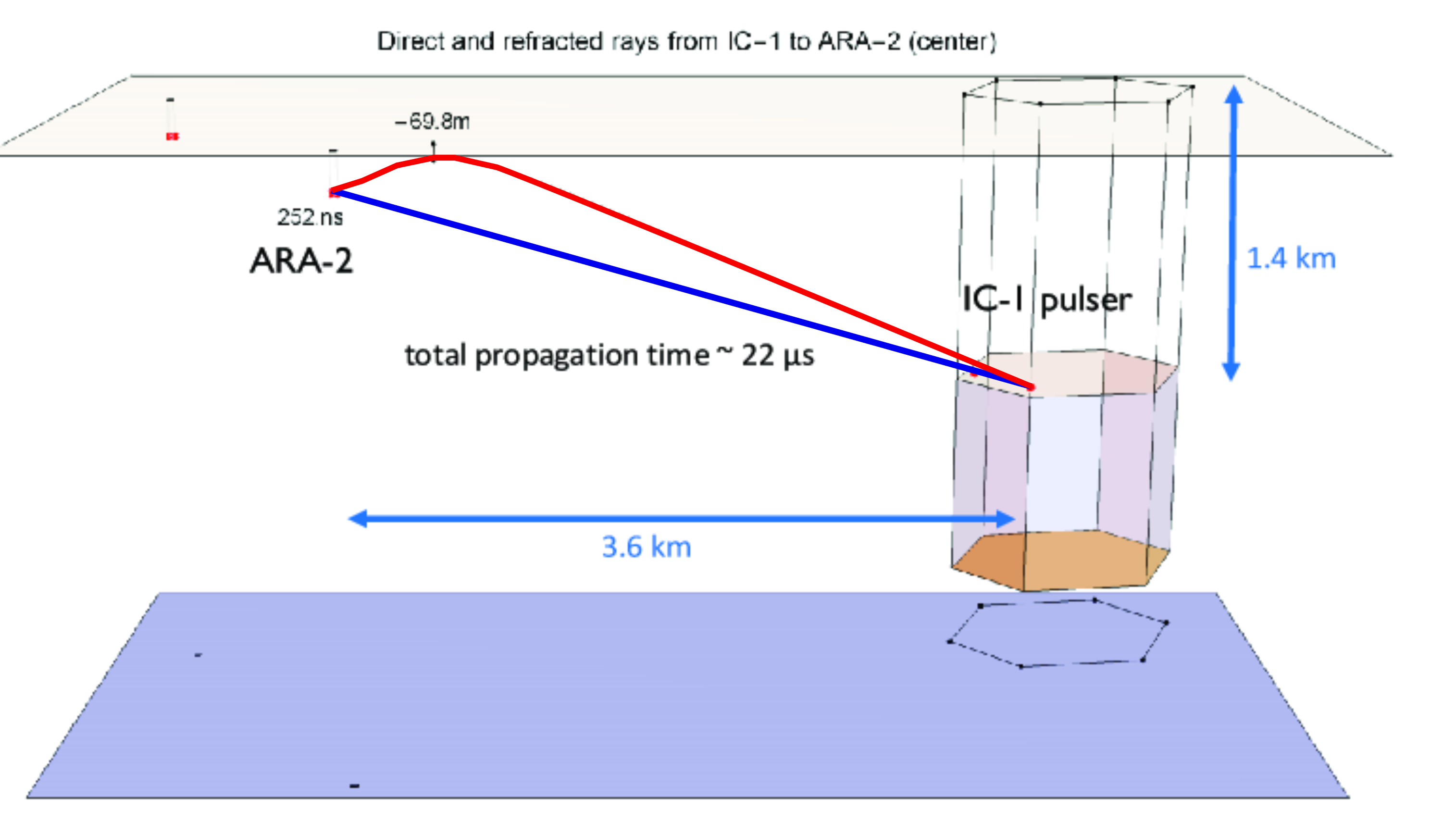}\caption{Trajectories of Direct and Refracted rays emitted from IceCube pulser IC1S and propagating to ARA Station 2. In this case, there is a 252 ns lag between the signal arrival at the 200-m deep ARA-2 receiver of the Refracted ray (red), which inflects at a depth of --69.8 m, relative to the Direct ray (blue). Additional dimensions/times are as shown in the Figure.}\label{fig:raytrace}\end{figure}

Several permittivity measurements are accessible to ARA.
Comparing the arrival time of received signals in the horizontally-polarized (HPol) to the vertically-polarized (VPol) receivers in an ARA station allows one to quantify possible polarization-dependence of electromagnetic wavespeed, aka `birefringence'. Comparison of the relative signal strength observed in a near receiver station with a far receiver station permits extraction of the absorption, per unit length, of radio waves in ice, corresponding to the imaginary portion of the dielectric permittivity. Owing to the variable index-of-refraction with depth in the ice, RF rays will follow arcs between source and receiver, with, in general, both a direct (`D') as well as a refracted (`R') path allowed, as illustrated in Figure \ref{fig:raytrace}. (If the signal geometry is sufficiently vertical, the refracted ray reaches the surface before `turning over' and is, in that case, more properly a reflected ray.)
\begin{figure}[htpb]\includegraphics[angle=0,width=0.9\textwidth]{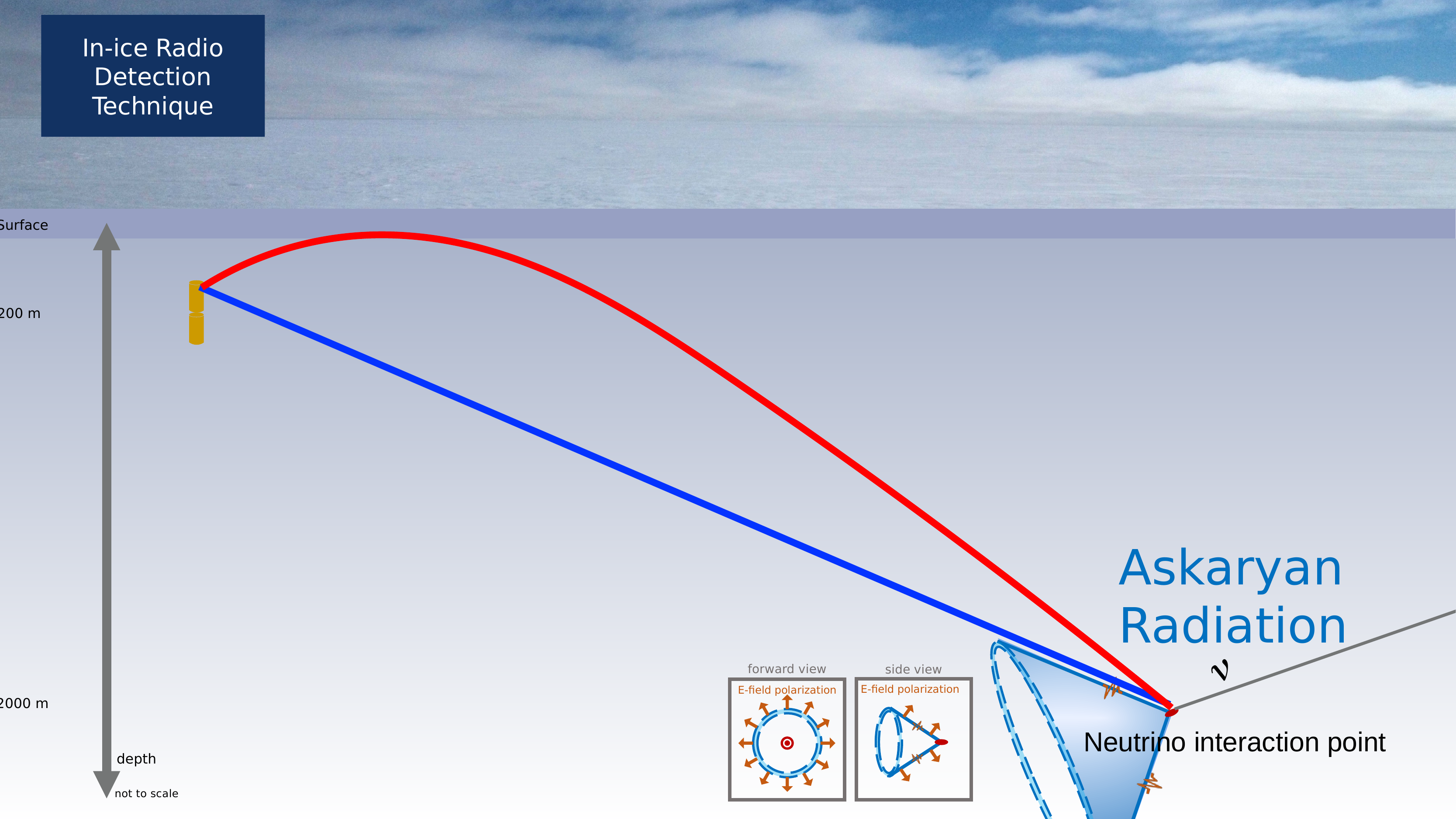}\caption{Modeled trajectories of Direct and Refracted rays emanating from the vertex of a
Cherenkov cone, following an
in-ice neutrino interaction. Note the similarity to the calibration signals illustrated in the previous Figure.}\label{fig:AskaryanDiagram}\end{figure}
Such a time-delayed 'double pulse' signal constitutes an unambiguous signature for an in-ice neutrino interaction. Figure \ref{fig:AskaryanDiagram} illustrates the geometric similarity
between the time-difference calibration measurements described herein and the sought-after neutrino signal. 

If the horizontal range is known {\it a priori},
the 100-500 ns separation in arrival time between the D and R signals can also be used to discriminate between models for the refractive index variation {\tt n(z)} with depth into the South Polar ice sheet. Alternatively, if the dependence of the time difference on range is known, a measured time difference can be used to infer the range to an in-ice source\cite{allison2019measurement}. A smooth depth variation of {\tt n(z)} also has an important consequence, namely, that there is a 'shadow zone' for which there are no solutions to Maxwell's Equations when both receiver and transmitter are located near, and slightly below the ice-air interface. In the Huygens picture, shadowing corresponds to a complete cancellation of all signal wavelets in this region. The extent of this shadow zone depends on both the lateral displacement of Tx and Rx, as well as their depths. For a typical ARA receiver depth of 180 meters, and a lateral displacement of 2 km (typical of the measurements presented herein), the transmitter shadow zone extends downwards from the surface to approximately 500 meters transmitter depth. Recently, and somewhat unexpectedly, propagation of signals from within the shadow zone have been reported\cite{barwick2018observation,deaconu2018measurements}, possibly linked to local (meter-scale) variations in the index-of-refraction profile and/or internal layering. As shown below, our data shows evidence for additional anomalies in radio-frequency propagation through South Polar ice.

\subsection{SPICE core transmissions}
We now outline the geometry and configuration for data-taking. 
\subsubsection{Ice cores and the SPICE core}
The original glaciological goal of the South Pole Ice Core Experiment (SPICE)\cite{casey20141500} was to measure, and also track changes in atmospheric chemistry, climate, and biogeochemistry over time, based on the measured properties of extracted cores. The SPICE core at South Pole is one of the few cores from the polar plateau and therefore fills in essential information between the Vostok\cite{petit1999climate}, Dome C\cite{lambert2008dust} and WAIS divide cores\cite{fitzpatrick2014physical}. Measurements at $\sim$meter scales offer the opportunity to link macroscopic studies of RF propagation with the microscopic properties of ice crystals. On the molecular scale, ${\rm H_2O}$ molecules in the familiar ice-Ih phase self-order into a 6-fold rotationally symmetric lattice, with oxygen atoms at each of the planar hexagonal vertices. Locally, cm-scale ice grains are typically parameterized in terms of a 3-dimensional `c'-axis, perpendicular to the hexagonal plane comprised by six ice molecules. The global orientation of the crystal (``COF'', or Crystal Orientation Fabric) is generally referenced relative to vertical (by convention, defined as ${\hat z}$) and a horizontal direction -- although coring devices thus far do not usually record azimuthal orientation, a reasonable ansatz aligns the x-axis with the local ice flow direction, which is generally known, and, aside from gravity, the only other symmetry-breaking stress in the ice. With those assumptions, one can then quantify the projections of the ice crystal orientation onto the $(x,y,z)$ axes for each grain. Qualitatively, analysis of samples taken from multiple cores leads to a standard picture in which the c-axes increasingly rotate towards compressional axes (in this case, gravity) with depth. Quantifying the c-axis alignment in terms of ellipsoidal principal moments $<{\hat e}_1>$ (parallel to ice flow), $<{\hat e}_2>$ (perpendicular to ice flow) and $<{\hat e}_3>$ (vertical, and coincident with ${\hat z}$), then, at ice depths to $\sim$500 meters, the ice fabric is largely randomly aligned ($<{\hat e_1}>\approx<{\hat e_2}>\approx<{\hat e}_3>$). Over the next kilometer in depth, the locus of c-axes forms a `girdle' through which the ice `flows' as compressional effects become increasingly noticeable and the c-axis distribution favors the yz-plane
($<{\hat e_1}>\approx 0; <{\hat e_2}>\approx<{\hat e}_3>$). At greater depths, the vertical gravitational stress increasingly rotates the c-axes to point directly along $\pm$z ($<{\hat e_1}>\approx<{\hat e_2}>\approx 0; <{\hat e}_3>\approx 1$). The birefringent asymmetry for flow parallel vs. perpendicular to ice flow is therefore roughly proportional to the asymmetry between $<{\hat e}_1>$ and $<{\hat e}_2>$. Note that the principal moments represent the eigenvectors of the orientation axes; the distribution of values around those eigenvectors is typically presented in the form of a Schmidt diagram\cite{rigsby1960crystal}.

\subsubsection{South Pole University of Kansas Pressure Vessel Antenna (SPUNK PVA)}
\begin{figure}[htpb]\includegraphics[width=0.9\textwidth]{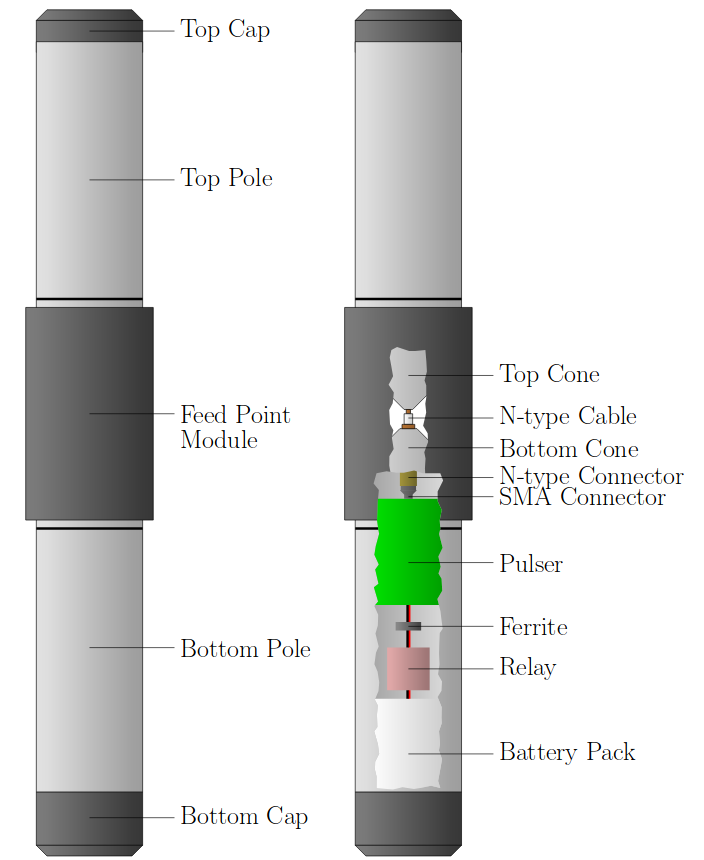}\caption{Pressure Vessel Antenna (aka ``SPICE Pulser from UNiversity of Kansas'', or SPUNK) used for broadcasts from the SPICE borehole. Vertical $\times$ horizontal dimensions of PVA are approximately 90 $\times$ 9 cm. Indicated in the figure are: i) Battery Pack used to power the ii) piezo, IDL or HVSP Pulser (in green), iii) Ferrite used to isolate Pulser from Relay, iv) Magnetically-sensitive relay (beige) allowing activation of unit after sealing, v) connection at feed point between Top and Bottom cones of dipole. Note that feed point is flooded with epoxy to stand off penetration from drilling fluid after immersion in SPICE core hole.}\label{fig:PVA}\end{figure}

Design of the SPUNK transmitter antenna was tailored to match the 
97-mm diameter of the SPICE core borehole, 
and the pressure of the estisol-240
drilling fluid environment into which
the antenna was immersed. An aluminum fat-dipole antenna design based on the original
RICE experimental model\cite{KravchenkoFrichterSeckel2003}, with one of the 
two cylindrical halves also 
serving as a sealed pressure vessel and containing batteries and a custom, high-voltage transmitter, was employed. 
As expected, the antenna broadcast efficiency improves when
fully immersed in the estisol-240 (n$\sim$1.42), as the intrinsic impedance of the dipole ($\sim$72$\Omega$ for an ideal dipole in vacuum) 
approaches the 50$\Omega$ impedance of the cable used to convey the transmitter signal
output to the antenna feedpoint. To ensure hermiticity, the feedpoint was itself flooded with epoxy to stand off the ambient
hydrostatic pressure. A diagram of the 
SPUNK PVA, with a cut-away, is shown in Figure \ref{fig:PVA}.

\section{Data-taking}
\subsection{Set-up}
\begin{figure}[htpb]
\centerline{\includegraphics[width=0.9\textwidth]{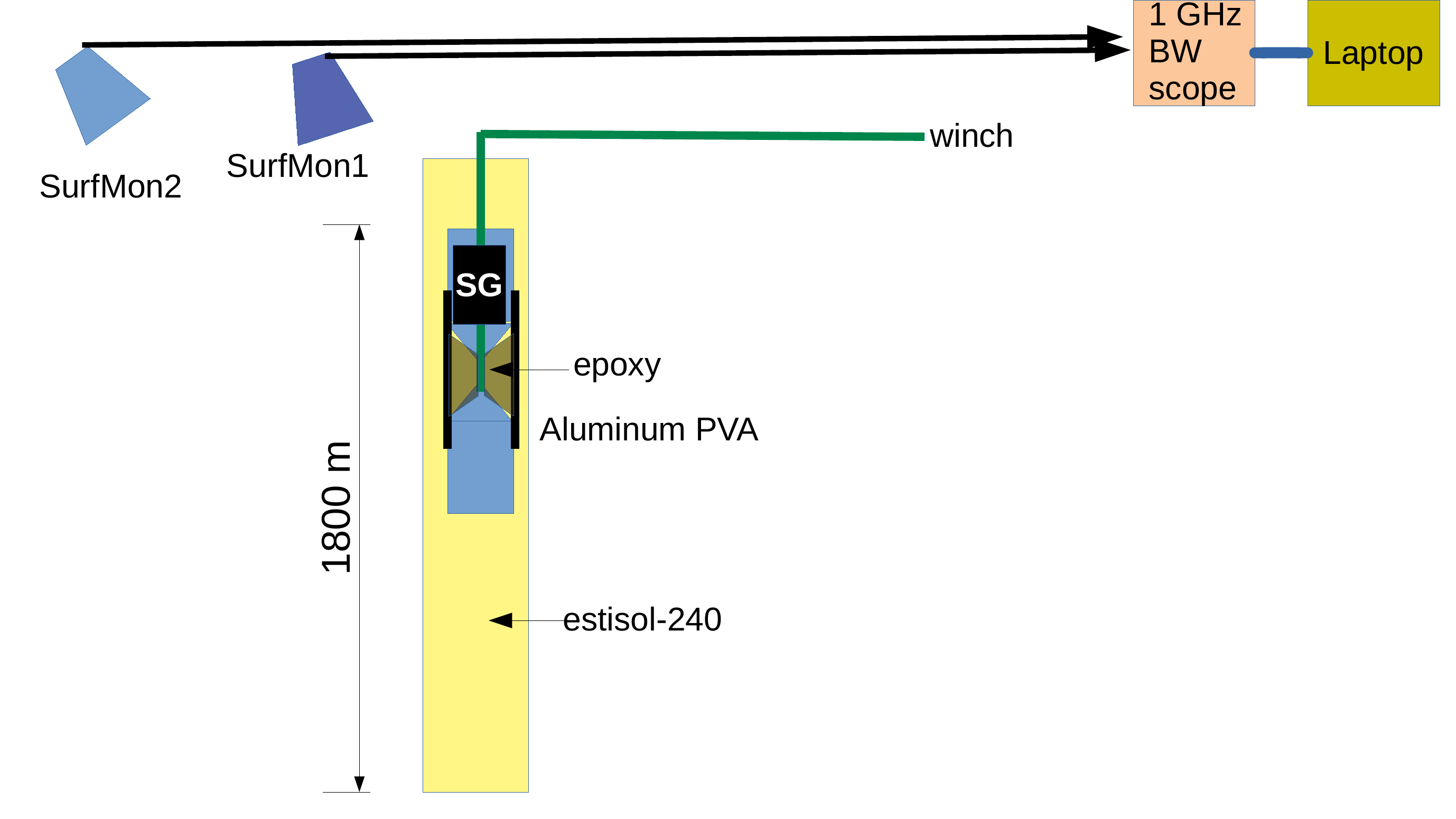}}
\caption{SPICE core pulsing configuration, showing the Pressure Vessel Antenna (PVA) immersed in estisol. Two surface horn antennas were used to verify real-time pulsing of the in-ice SPUNK PVA. Indicated in the Figure is the Signal Generator unit (comprising Pulser, Ferrite, Relay and Battery Pack from previous Figure). Winch controls only ascent/descent of PVA into estisol-240 drilling fluid, and has no power/communications functionality. To monitor Signal Generator in real-time, two surface monitor horn antennas (SurfMon1 and SurfMon2) are placed on surface approximately 5 meters and 50 meters, respectively, from SPICE core hole. Surface monitor
signals are conveyed by LMR-400 cable to a battery-powered 1 GigaHertz bandwidth Agilent digital sampling oscilloscope, with data from the oscilloscope written to laptop.}
\label{fig:spicegeo}
\end{figure}

\begin{figure}[htpb]
\centerline{\includegraphics[width=0.9\textwidth]{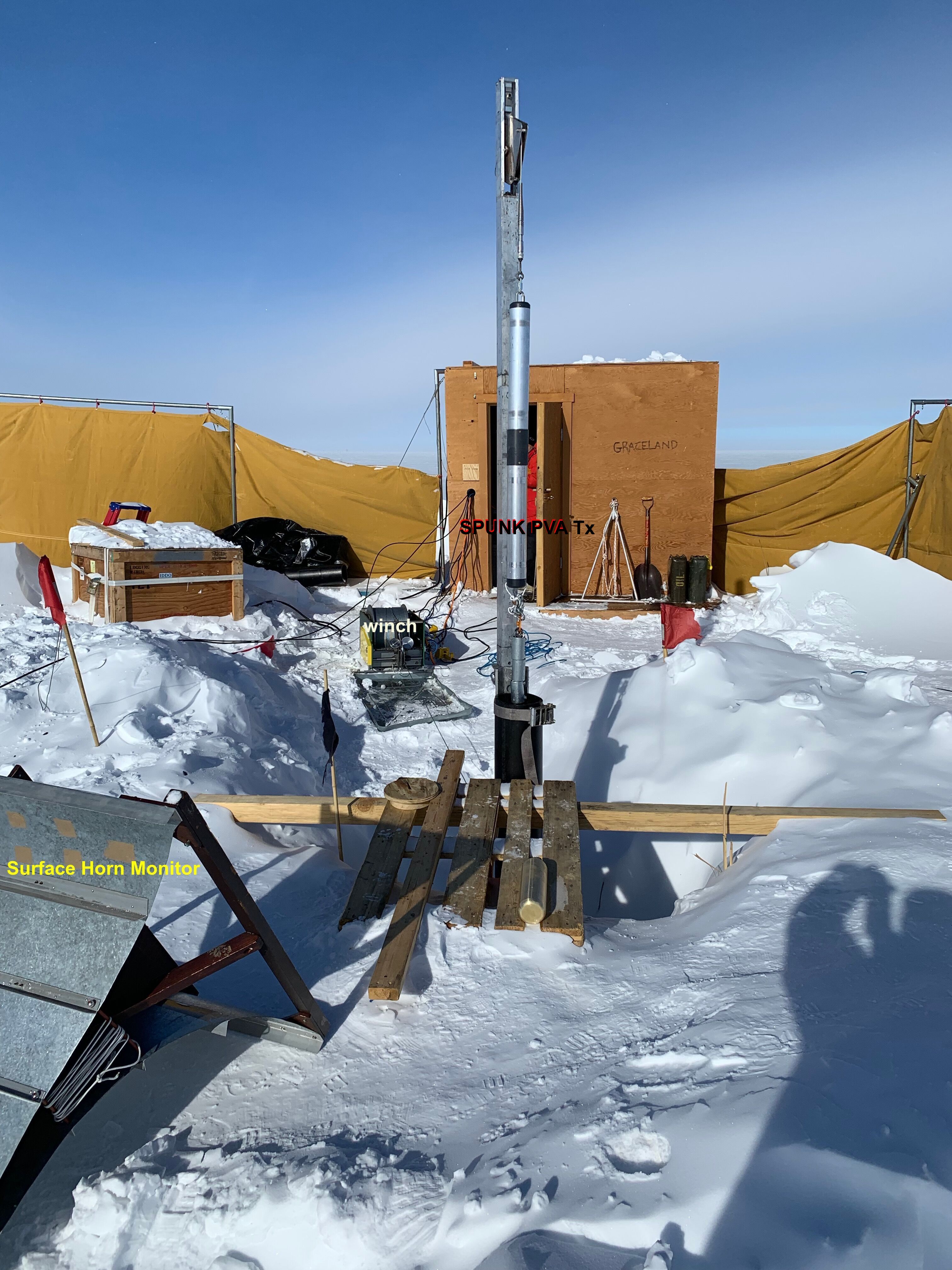}}
\caption{Photograph of experimental set-up prior to lowering transmitter antenna into hole. Visible are
vertically-aligned SPUNK PVA (indicated with black label), 
one of the two surface horn antenna monitors (SurfMon1, as indicated with yellow label) and 
winch (as indicated with white label). Note weight stack at the bottom of SPUNK PVA to ensure buoyancy after descent into
SPICE core hole, slightly obscured by above-surface black cylindrical hole casing.}
\label{fig:spicepic}
\end{figure}

Figure \ref{fig:spicegeo} illustrates the experimental layout during the eight day period (12/23/18--12/31/18, corresponding to Julian Days 358--365) over which SPUNK PVA
pulsing occurred. 
Two surface horn antennas
(designed and built by the Institute of Nuclear Research, Moscow) provided real-time monitoring on a surface oscilloscope of the emitted signals, and ensured that the pulser was active.
Figure \ref{fig:spicepic} shows a photograph of the experiment, just prior to lowering the SPUNK PVA into the 
SPICE core borehole.

\subsection{Transmitter$\to$Receiver Geometry}
\begin{figure}[htpb]
\centerline{\includegraphics[width=0.9\textwidth]{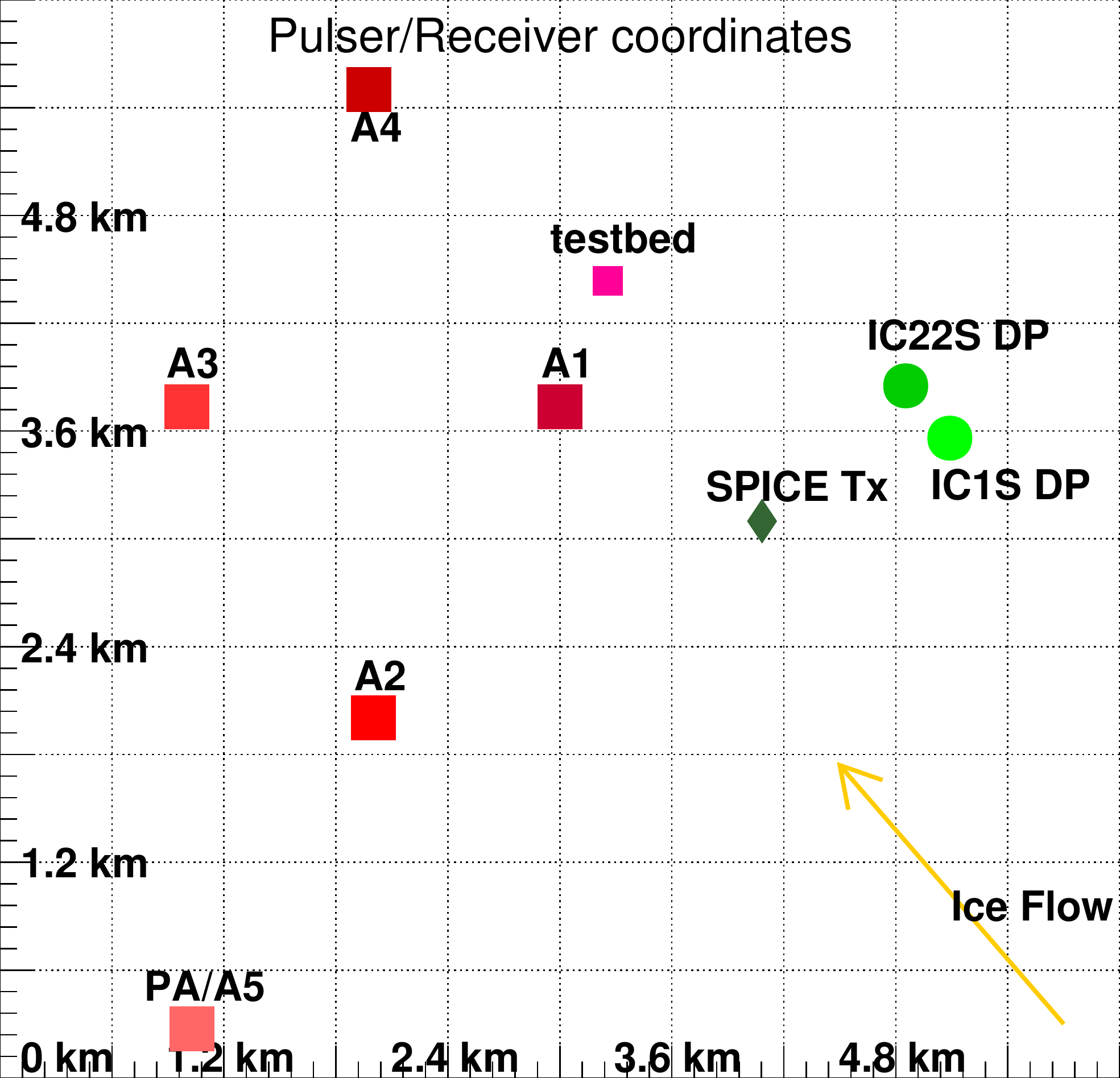}}
\caption{Geometry of ARA receiver stations (red/pink) relative to transmitter sources at IC1S, IC22S and SPICE borehole. Phased Array receiver string is co-located with
station A5. }
\label{fig:geom}
\end{figure}

Since birefringence is linked to the local crystal orientation,  
the relative alignment of the receiver stations with respect to the transmitter is important for interpreting results.
The geometry of the SPICE core (surveyed in the 2017-18 field season at 89.9889 S, 98.1596 W [or $89^o$ 59' 20.168'' S, $98^o$ 09' 34.528'' W]\footnote{We note that the local ice flow has magnitude $\sim$10 m per year, so the hole must be periodically re-surveyed, although the location relative to the ARA stations, which are co-moving, should remain approximately constant.} 
relative to the ARA stations is presented in 
Table \ref{tab:geom} as well as Figure \ref{fig:geom}, including the distances {\tt d} in meters between the centroid of a given ARA station (denoted ``A1''--``A5'', with the exception of the initial Testbed, as shown)
and the transmitter source (measured relative to either the top of the SPICE core, or one of the deep pulsers co-deployed with the IceCube array in 2011-12, to a depth of 1400 m, 
on either IceCube String 1 or String 22). Vis-a-vis the later discussion of birefringence, the azimuthal
angle $\phi$ between the line connecting the receiver to a given transmitter source and the local ice flow direction is also shown.

\begin{table}[htpb]
\begin{tabular}{c|c|c|c} 
        &          SPICE          &        IC1S          & IC22S \\
Station &  ${\tt d_{source}}$/$\phi_{source}$/$\phi_{ice~flow}$ & ${\tt d_{source}}$/$\phi_{source}$/$\phi_{ice~flow}$ & ${\tt d_{source}}$/$\phi_{source}$/$\phi_{ice~flow}$ \\ \hline
Testbed & (no data) & 2463 m/-25$^o$/23$^o$ & 2199 m/-20$^o$/29$^o$ \\
A1 & 1257 m/$-30^o$/18$^o$ & 2517 m/-4$^o$/44$^o$ & 2322 m/3$^o$/53$^o$ \\
A2 & 2353 m/28$^o$/76$^o$ & 3729 m/27$^o$/76$^o$ & 3674 m/33$^o$/82$^o$ \\
A3 & 3146 m/-12$^o$/37$^o$ & 4323 m/-2$^o$/47$^o$ & 4099 m/2$^o$/50$^o$ \\ 
A4 & 3199 m/-49$^o$/1$^o$ & 3865 m/-32$^o$/17$^o$ & 3534 m/-30$^o$/19$^o$ \\
A5 & 4164 m/43$^o$/91$^o$ & 5364 m/39$^o$/88$^o$ & 5377 m/43$^o$/92$^o$ \\ \hline
\end{tabular}
\caption{Lateral distances and azimuthal orientation relative to both the indicated sources, as well as local ice flow, with $\phi_{ice~flow}$=0 defined as parallel to the local ice flow direction, for measurements described herein. We note that two of the ARA stations (A2 and A5) are oriented predominantly perpendicular to the local ice flow direction, while stations A1 and A4 are oriented predominantly parallel to the ice flow direction. Typical errors are estimated to be approximately one degree.}
\label{tab:geom}
\end{table}
The most detailed data at South Pole on the ice flow is provided by the IceCube experiment itself, who have measured the velocity field over a roughly one square kilometer scale. 
Local angular deviations from uniform flow of several degrees, over the $km^2$ measurement area covered, are not uncommon, and indicative of the angular systematic error for the values presented in Table \ref{tab:geom}.

\begin{figure}[htpb]
\centerline{\includegraphics[width=0.9\textwidth]{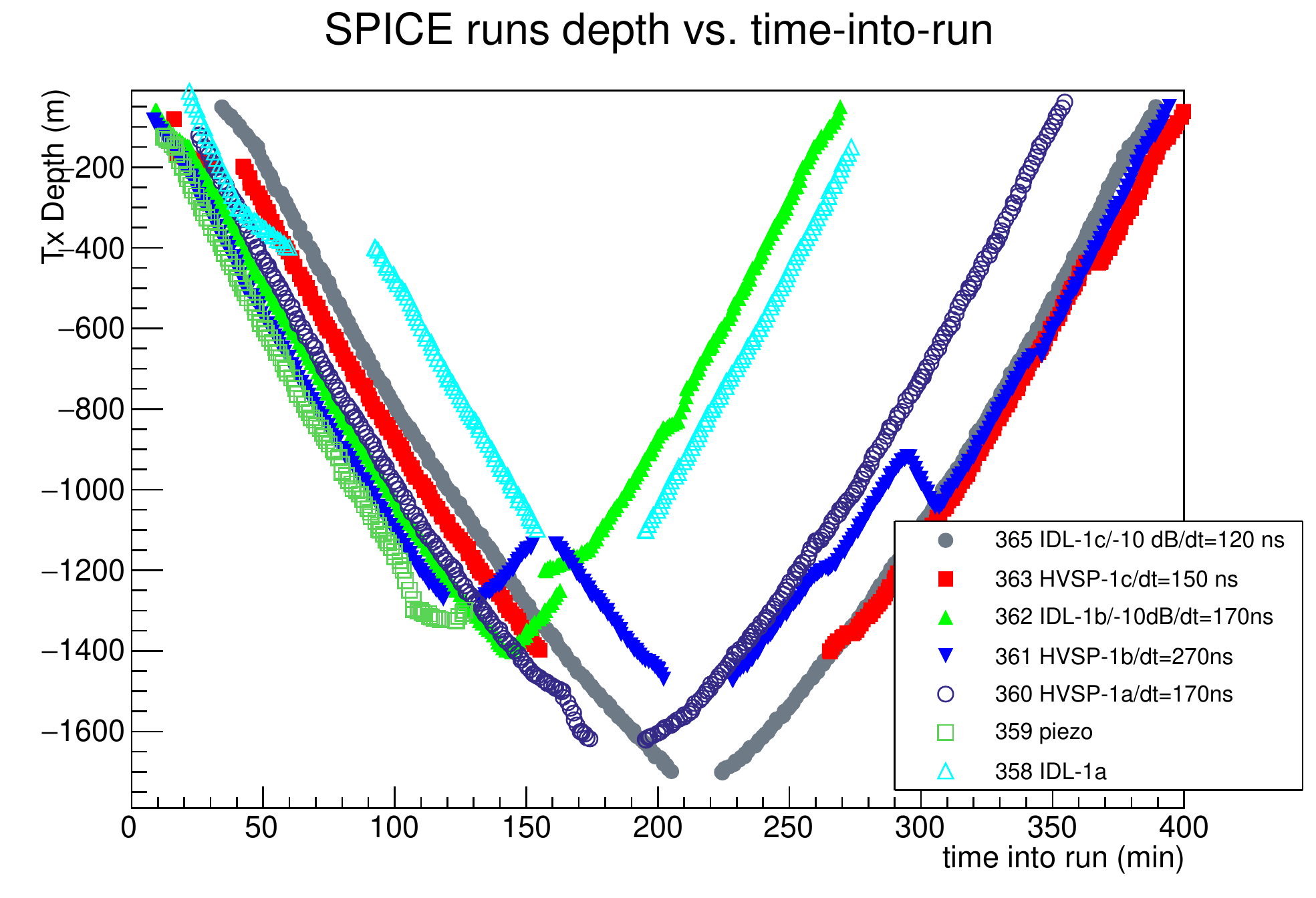}}
\caption{Summary of SPICE core SPUNK PVA drops, conducted from Dec. 23--Dec. 31, 2018. In all cases, t=0 corresponds to start of antenna drop. Also indicated are the Julian day of the run, the pulser used (piezo, IDL or HVSP), the attenuation used on the transmitter output (10 dB for runs on days 365 and 362), and the 3-fold coincidence window used for the ARA receiver stations trigger.}
\label{fig:DepthVTime}
\end{figure}

Over a period of
6--8 hours,\footnote{We are indebted to Ryan Bay, Summer Blot and Anna Pollmann, who spent several person-days to ensure that the 3 kW generator used for power, as well as the winch used for lowering SPUNK into the SPICE borehole were both in excellent working order when we arrived at South Pole. A platform constructed at the top of the hole also significantly facilitated daily set-up and take-down.} the transmitter antenna was lowered into the SPICE borehole to an 1100--1700 depth, and then retracted back to the surface. Figure \ref{fig:DepthVTime} illustrates the depth vs. time profile for the seven days over which data were taken. The three pulsers used include the piezo model used for the HiCal experiment\cite{gorham2017hical,gorham2017antarctic,prohira2018antarctic,prohira2019hical}, and transmitters based on generating high-voltage across a spark gap (``HVSP''), 
as well as a fast high-voltage discharge generated by a transistor array (``IDL''); both of the latter two models include fast
DC$\to$DC converters in their design.

\begin{figure}[htpb]
\centerline{\includegraphics[width=0.95\textwidth]{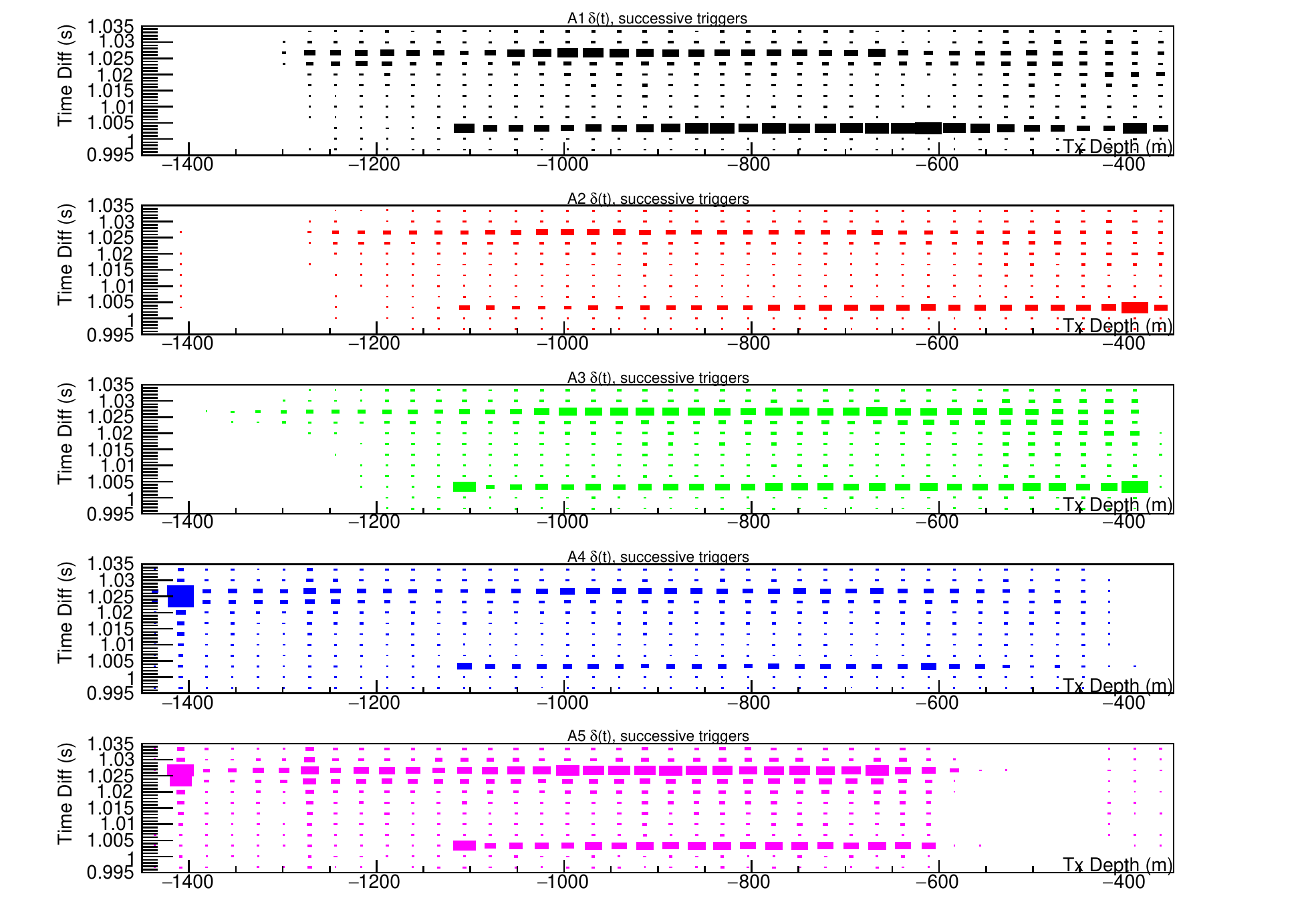}}
\caption{Time difference between successive triggers for ARA receiver stations (vertical axis, in seconds), as function of SPUNK depth (in meters), showing the $1028\pm0.9$ ms period of the HVSP (Julian days 360, 361 and 363) and $1004\pm 0.7$ ms period of the IDL (Julian days 358, 362 and 365) PVA transmitters. The A1--A3 trigger performed irregularly for depths greater than approximately 1200 m.}
\label{fig:dttrigs}
\end{figure}

Since the SPUNK transmitter within the PVA upper dipole chamber 
was powered by battery, after initial activation at the surface, the transmitter broadcast
continuously over the entirety of the transmitter drop. To facilitate offline identification of pulses
from the SPUNK pulser, with the exception of the mechanically-activated piezo-electric transmitter, which exhibited a temperature-dependent
pulser period between 3--8 seconds, and also emitted noticeable after-pulses, signals were emitted by the HVSP and IDL pulsers at approximately 1 Hz. Fig. 
\ref{fig:dttrigs} shows the time-interval between successive ARA station triggers during broadcasting. 
The periods of the two pulsers are
evident from Figure \ref{fig:dttrigs}. \message{We note that not all of the ARA station clocks were globally synchronized during this data-taking.}

\section{Observations and received signal waveforms}
\begin{figure}
\centerline{\includegraphics[width=0.95\textwidth]{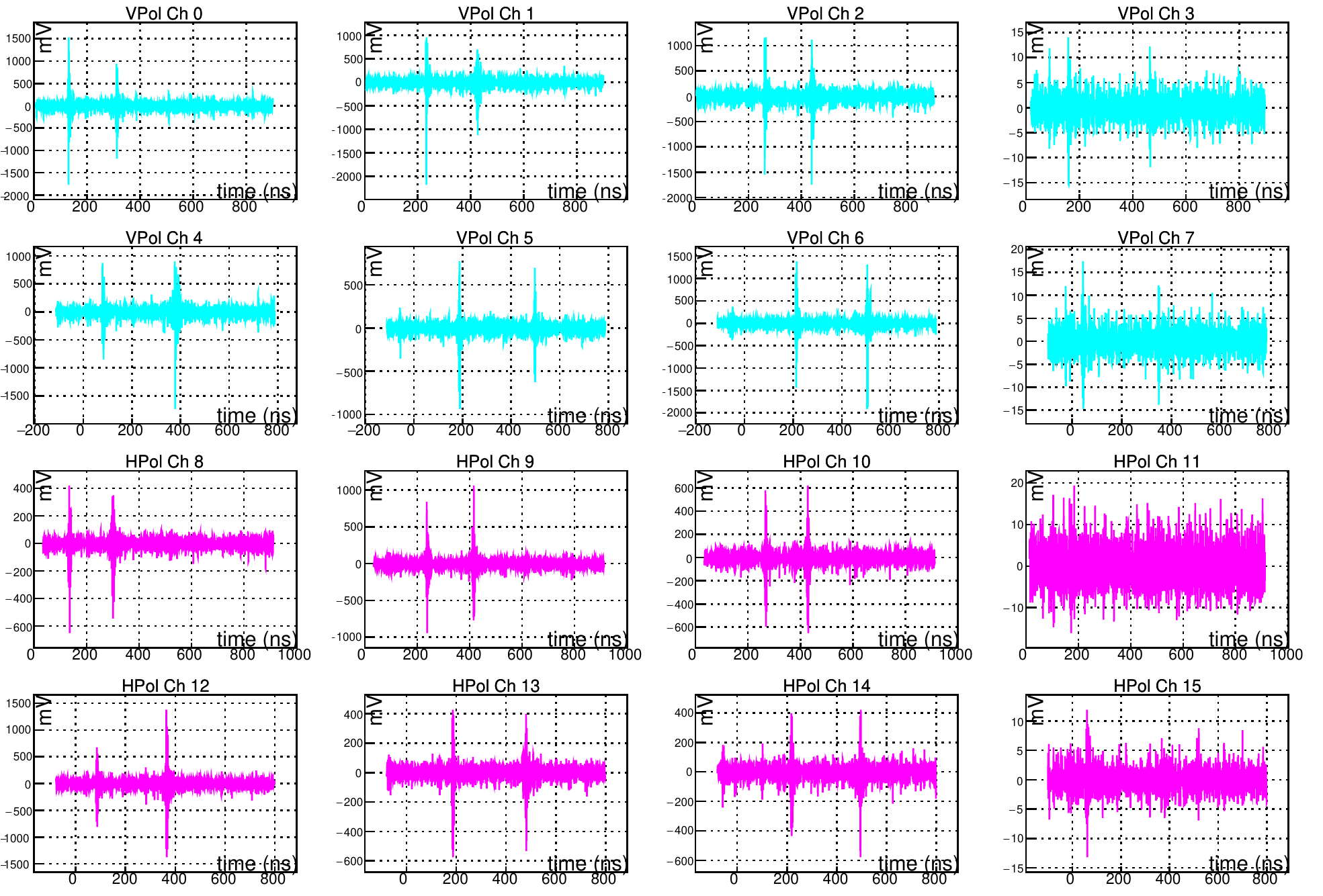}}
\caption{A4 event display (milliVolts [y-] vs. nanoseconds into waveform ([x-]) for four top VPol antennas (top row, cyan), bottom VPol antennas (second row, cyan), top HPol antennas (third row, magenta) and bottom HPol (fourth row, magenta) antennas during SPUNK PVA pulsing. During this data-taking the gain for string 4 (VPol Ch3, VPol Ch7, HPol Ch 11, and HPol Ch 15) was reduced relative to other channels.}\label{fig:A4evdisp}\end{figure}

Figure \ref{fig:A4evdisp} shows a waveform display for an event trigger recorded in ARA station A4. In typical events, approximately 1000 ns of data
are recorded at a sample rate of 3.2 GSa/sec.
In the Figure, the top two rows correspond to the voltage vs. time profiles for the VPol channels; the lower two rows present the voltage vs. time for the HPol channels. Conspicuous in these plots are `double-pulses', one resulting from the direct trace ('D') from source to receiver, and a second, delayed signal resulting from refraction ('R') through the ice, characteristic of broadcasts from the `visible' zone, and below the shadow boundary. Within the data acquisition system (DAQ), the channels are ganged according to the icehole (`string') into which the antennas are deployed. During the time that these data were taken, the gain on the channels in string 4 was noticeably reduced. 

\begin{figure}[htpb]  \includegraphics[width=\textwidth]{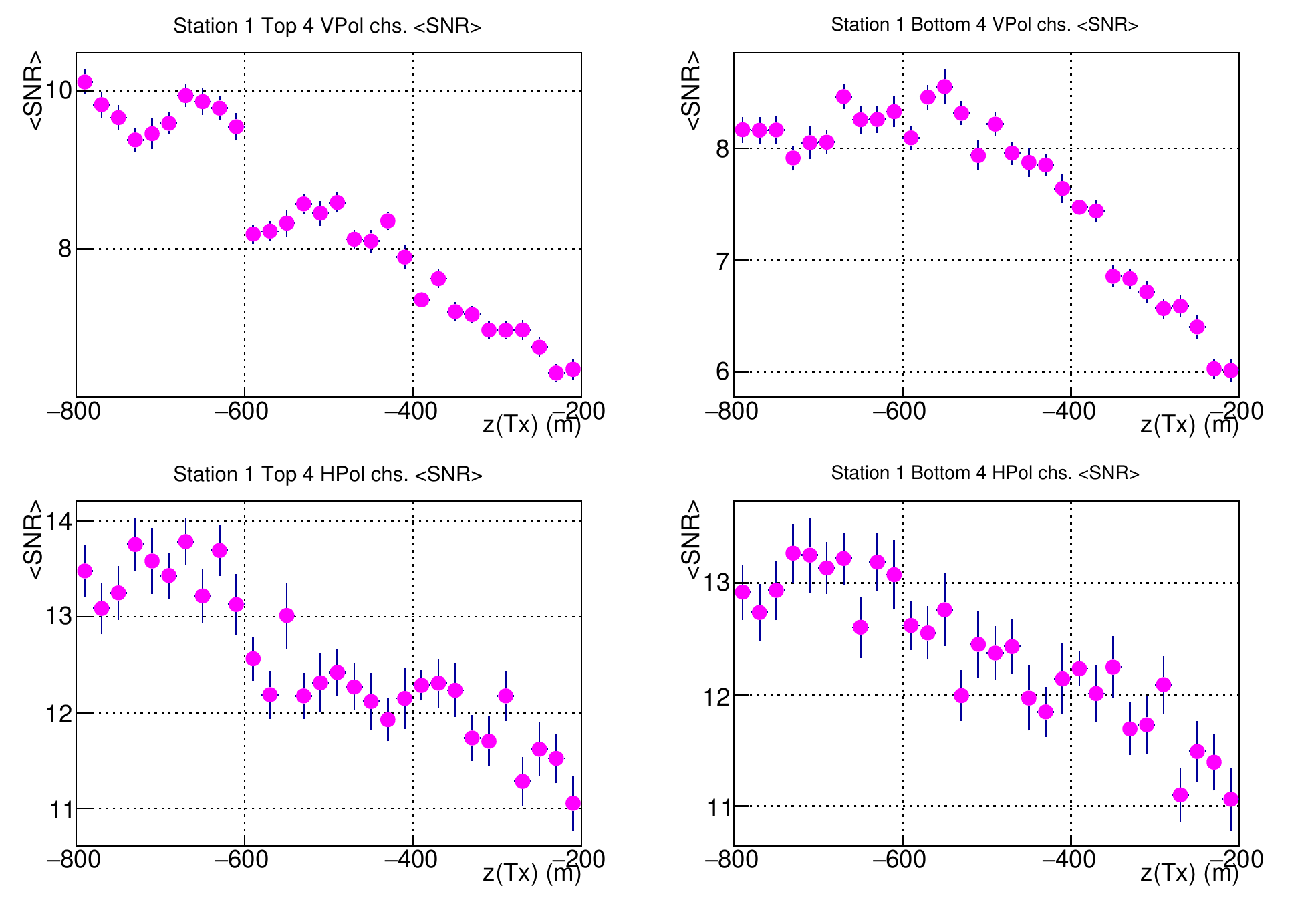} \caption{A1 Signal-to-Noise ratio, as function of transmitter depth. Note the
enhancement in VPol SNR as the transmitter crosses the shadow zone boundary at approximately z=-600 m.}\label{fig:A1SNR}\end{figure}

\begin{figure}[htpb]  \includegraphics[width=\textwidth]{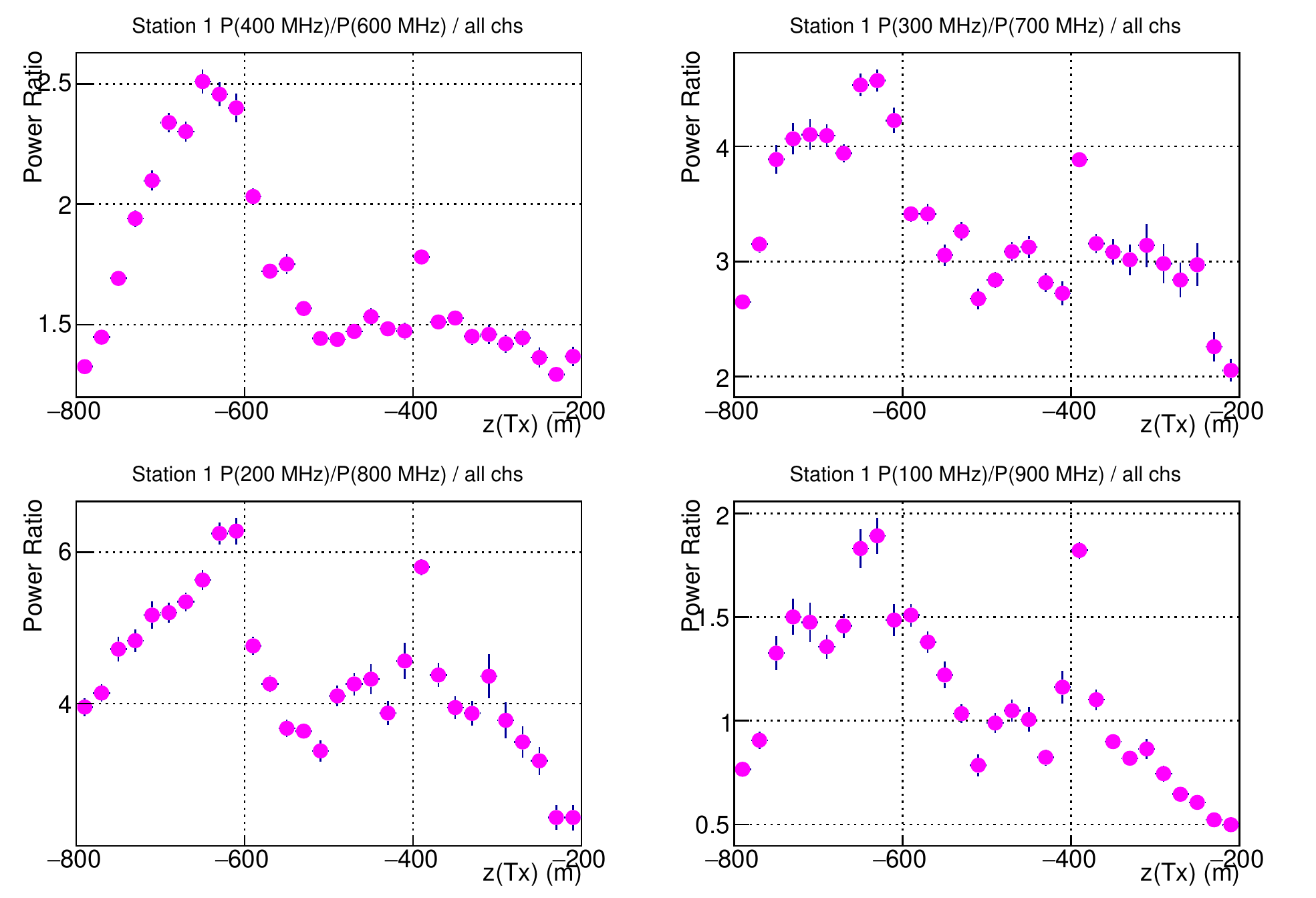} \caption{A1 ratio of VPol Signal Power in indicated frequency bands, as function of Tx depth. Top left panel, e.g., displays total signal power in 350 MHz -- 450 MHz band relative to 550 MHz -- 650 MHz band.}\label{fig:A1FFT}\end{figure}

\begin{figure}[htpb]  \includegraphics[width=\textwidth]{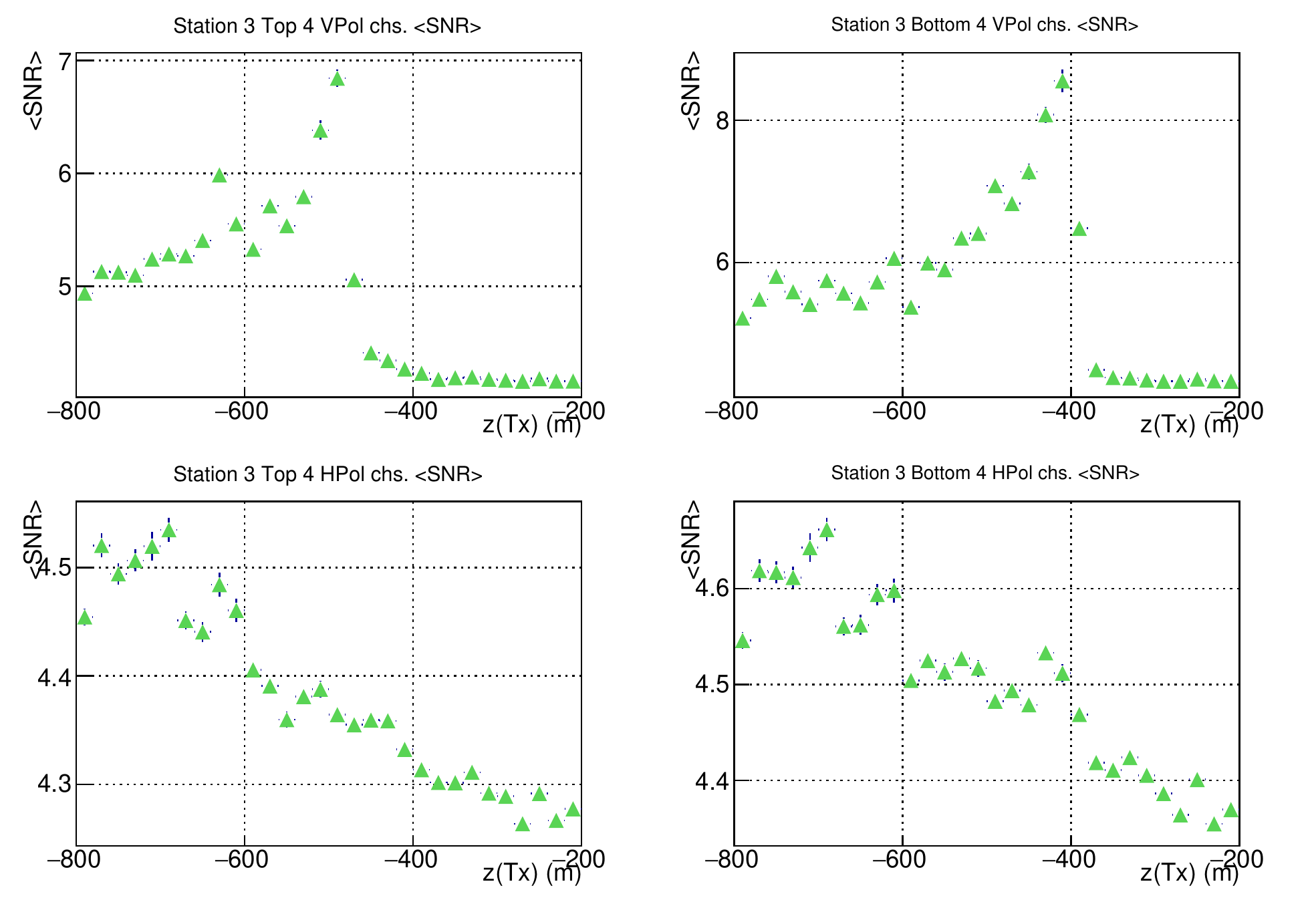} \caption{A3 Signal-to-Noise ratio, as function of transmitter depth.}\label{fig:ASNR}\end{figure}

\begin{figure}[htpb]
\includegraphics[width=0.65\textwidth]{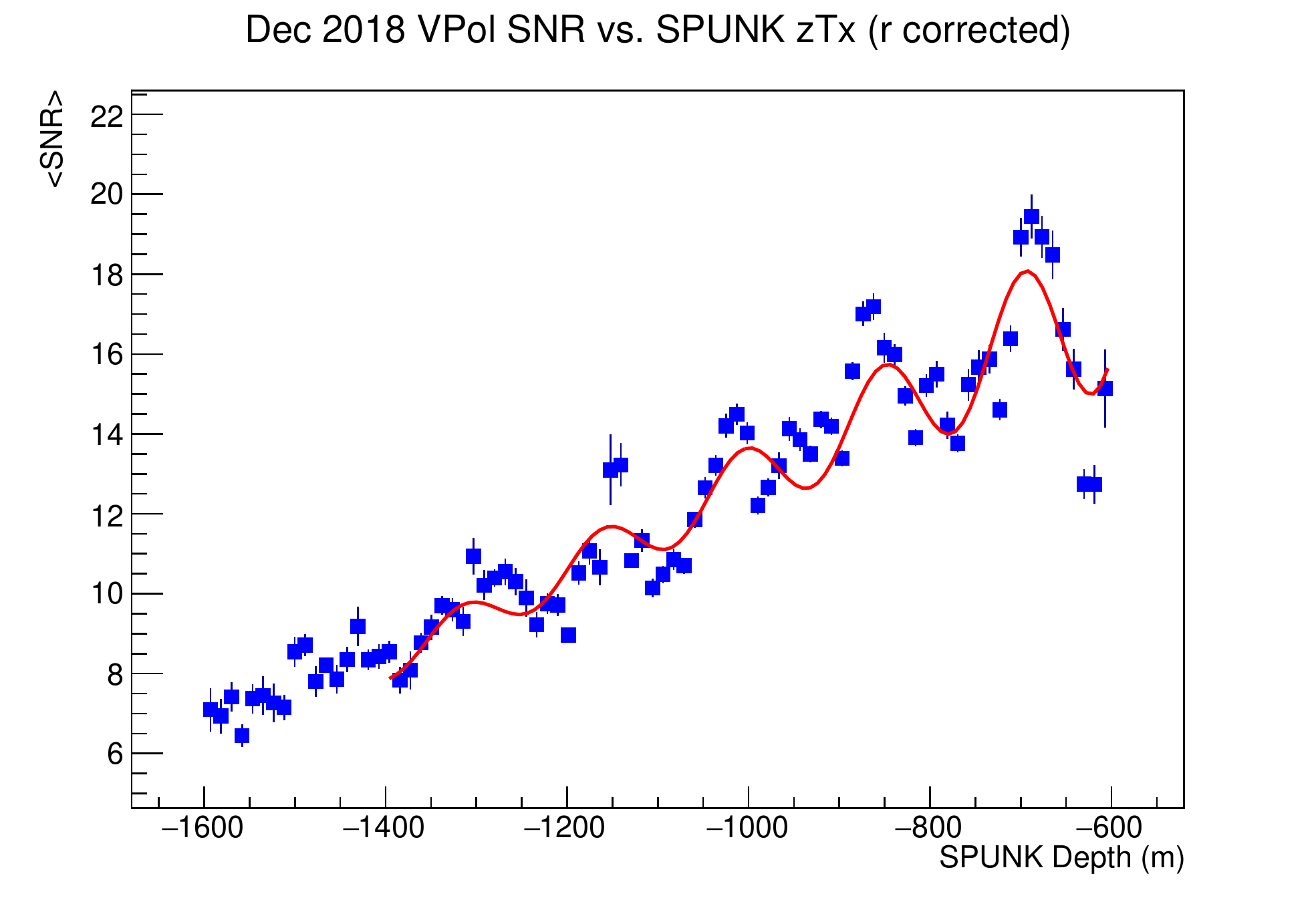}
\caption{VPol Station A5 measured Signal-to-Noise ratio as a function of depth.
The SNR data have been fit to a functional form
$A(z)=A_0\cos(kz+\phi_0)$.}
\label{fig:A5SNRvzTx}
\end{figure}

We note considerable signal power in the HPol channels (lower two rows) in Figure
\ref{fig:A4evdisp}. The cross-polarization SPUNK PVA signal strength in the lab, 
in pre-deployment testing, was observed to be reduced by 6 dB in power relative to
the co-polarization signal strength, typical of the fat dipole model. 
For the {\it in situ} SPICE tests, and assuming the
estisol+ice environment does not significantly change the SPUNK PVA beam,
the measured HPol:VPol signal strength at the receiver 
expected for the vertically-oriented SPUNK PVA transmitter should
be reduced by the product of that 6 dB transmitter cross-polarization suppression, multiplied by the intrinsic relative HPol:VPol receiver gain ratio over the spectral
bandwidth corresponding to the SPUNK PVA output signal. Empirically, we observe 
unexpectedly large variation in 
both the HPol, as well as (independently) the VPol signal strength as a function of depth into the SPICE
borehole, for both shallow (shadowed) as well as deeper (non-shadowed) geometries, suggestive of multi-path
interference effects. 
\message{Investigating this phenomenon in more detail will be one of the primary scientific objectives of the 2020-2021 SPICE core tests.}
Selecting the stations with the most vertical signal incidence angles (A1), intermediate elevation signal incidence angles (A3), and
the most horizontal incidence angles (A5), 
additional empirical features of the measured data are presented in Figures \ref{fig:A1SNR}--\ref{fig:A5SNRvzTx}. 
Particularly for A3,
we observe significant VPol signal enhancement around the depth of the shadowed/non-shadowed transition, as expected for
flux focusing around a caustic. 
The power spectra shown for A1 exhibit an evident shift to lower frequencies (f$<$500 MHz) as the transmitter emerges from the shadowed regime.
Figure \ref{fig:A5SNRvzTx} shows the A5 station
signal-to-noise ratio, as a function of transmitter depth. We expect a gradual reduction in observed signal owing to the warmer, more absorptive ice being sampled as the transmitter descends (1/r effects are negligible for this
geometry, and have been corrected for in the Figure.). 
We also observe an apparent modulation of the signal amplitude with depth; similar features have been reported 
in concurrently-recorded ARIANNA data\cite{anker2020probing}.
Although we cannot
rule out the possibility that this is an antenna effect, we note that no such angular-dependence in signal strength was observed in pre-deployment anechoic chamber calibration.
Overall, we note considerable deviations
from smoothness in these distributions.
Although a comprehensive first-principles model for shadow and near-shadow zone propagation has yet to be developed, a successful model should, in principle, reproduce these observations. 

\section{Hit-Finding and Pattern Recognition}
\begin{figure}[htpb]
\centerline{\includegraphics[width=0.95\textwidth]{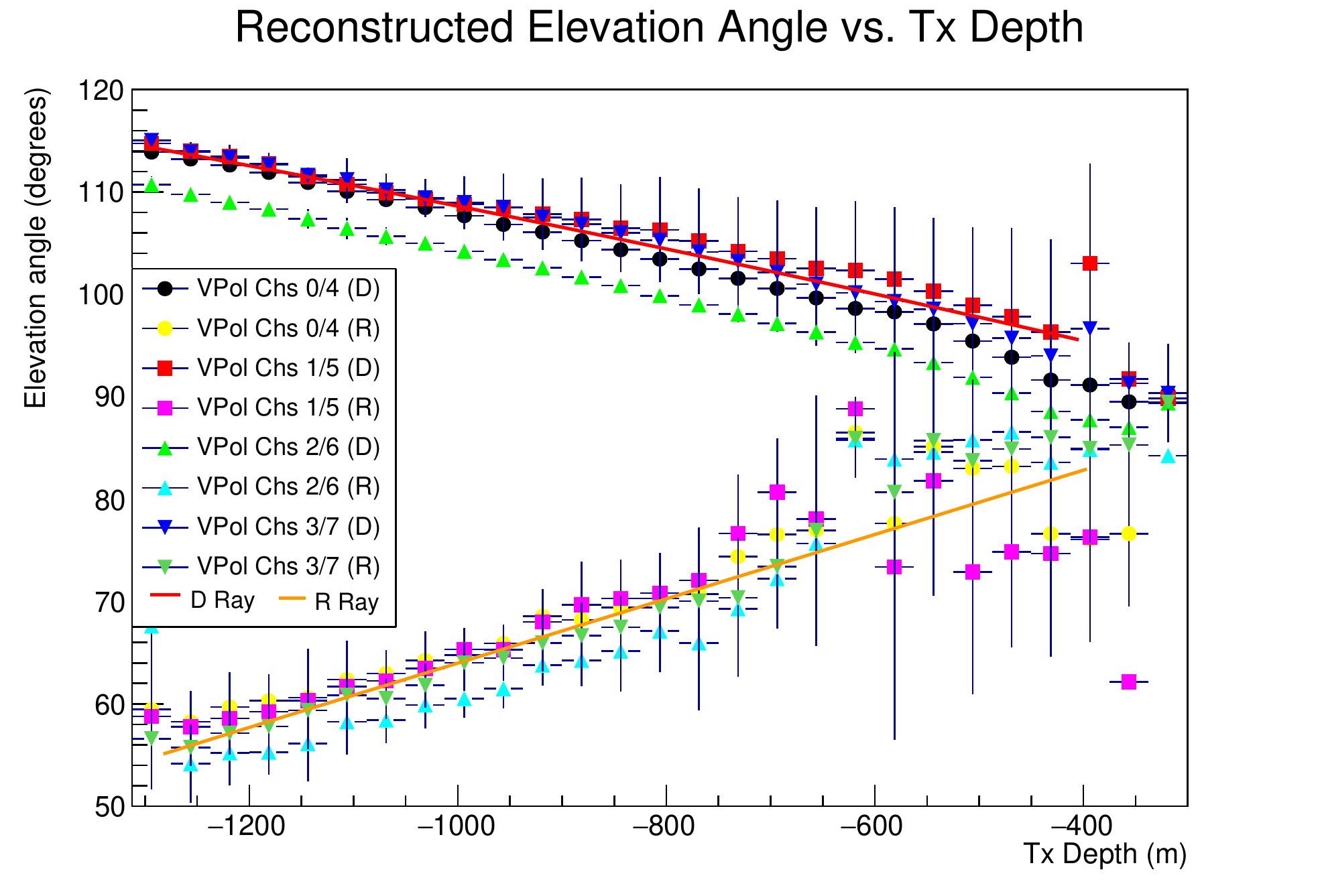}}
\caption{A2 reconstructed (points) incident elevation angles for Direct and Refracted rays, as a function of
depth of SPUNK PVA. Overlaid curve
shows expectations from ray tracing. Note that at shallow depths, the decreasing time difference, and therefore increasing overlap between the D and R rays impedes separating the two signals in a given waveform.}
\label{fig:Angledegcz}
\end{figure}

\begin{figure}[htpb]
\includegraphics[width=0.95\textwidth]{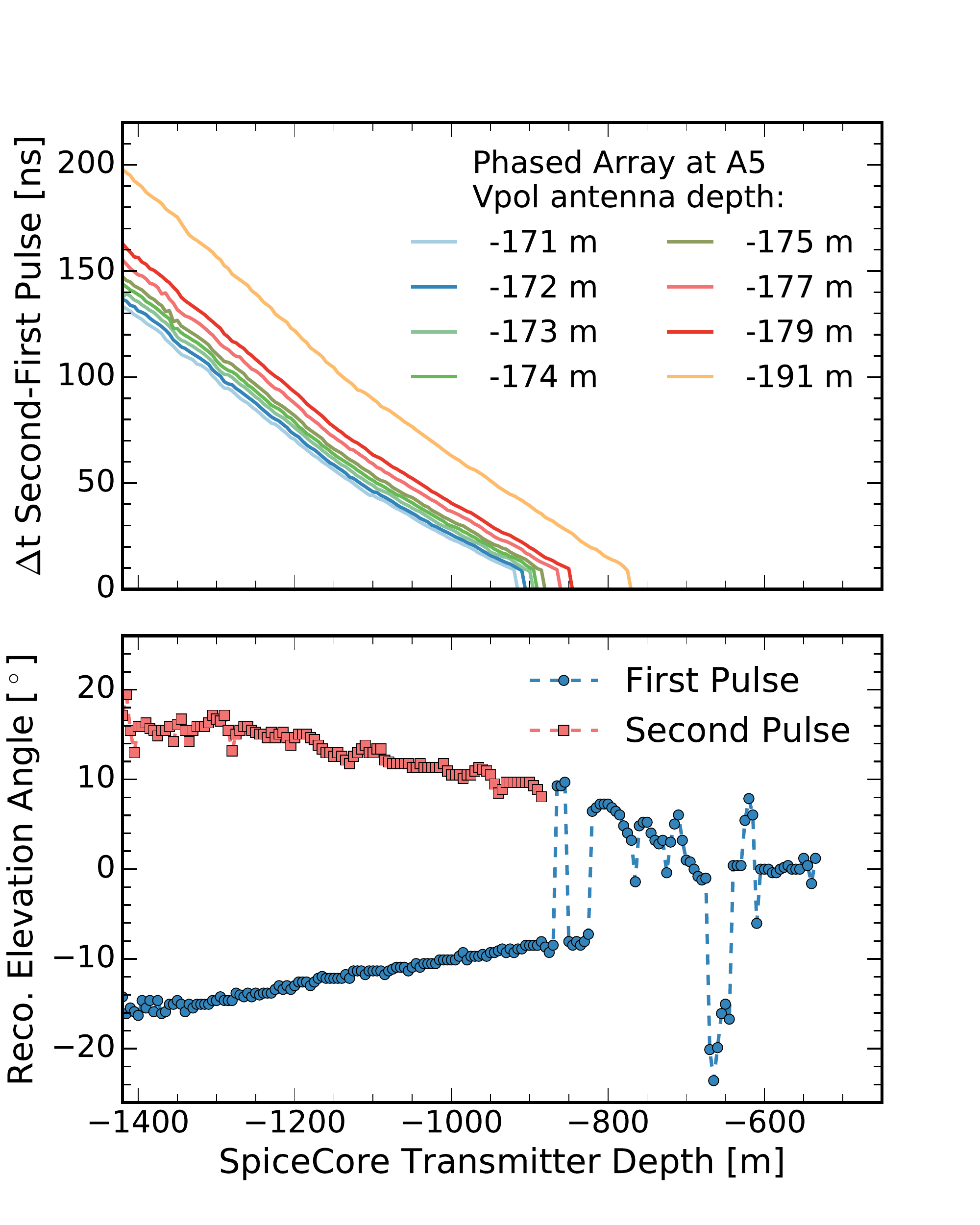}
\caption{(D,R) time difference (top) and also elevation reconstruction (bottom) for Phased Array.}
\label{fig:PA}
\end{figure}

The standard ARA source reconstruction algorithms utilize channel-to-channel waveform cross-correlations 
to extract relative signal arrival times in the antennas that comprise a single station. Combined with 
antenna geometry and location information, relative `hit' times on two vertically-displaced channels (i.e., one hit-time difference) are sufficient to infer 
a signal arrival direction in elevation, assuming an incident plane wave. 
As shown in Figure \ref{fig:A4evdisp}, there are doublets of pulses arriving within a typical $\cal{O}$(1000 ns) window --
the direct D signal arriving from below the centroid of the array, followed by the refracted R pulse arriving from above the array centroid.
In Figure \ref{fig:Angledegcz}, we show the inferred elevation angle for same-hole channels, by
projecting the incident plane wave onto the vertical (z-) axis. As expected, as the
transmitter descends into the SPICE core icehole, the D and R incidence angles deviate approximately symmetrically
from $90^\circ$.
Figure \ref{fig:PA} displays the corresponding data derived from the Phased Array\cite{vieregg2016technique}, deployed at the location of A5, and consisting of a single
string of multiple VPol antennas, qualitatively consistent with the features of Fig. \ref{fig:Angledegcz}.

\subsection{Direct and Refracted Double Pulses}
\begin{figure}[htpb]
\centerline{\includegraphics[width=0.95\textwidth]{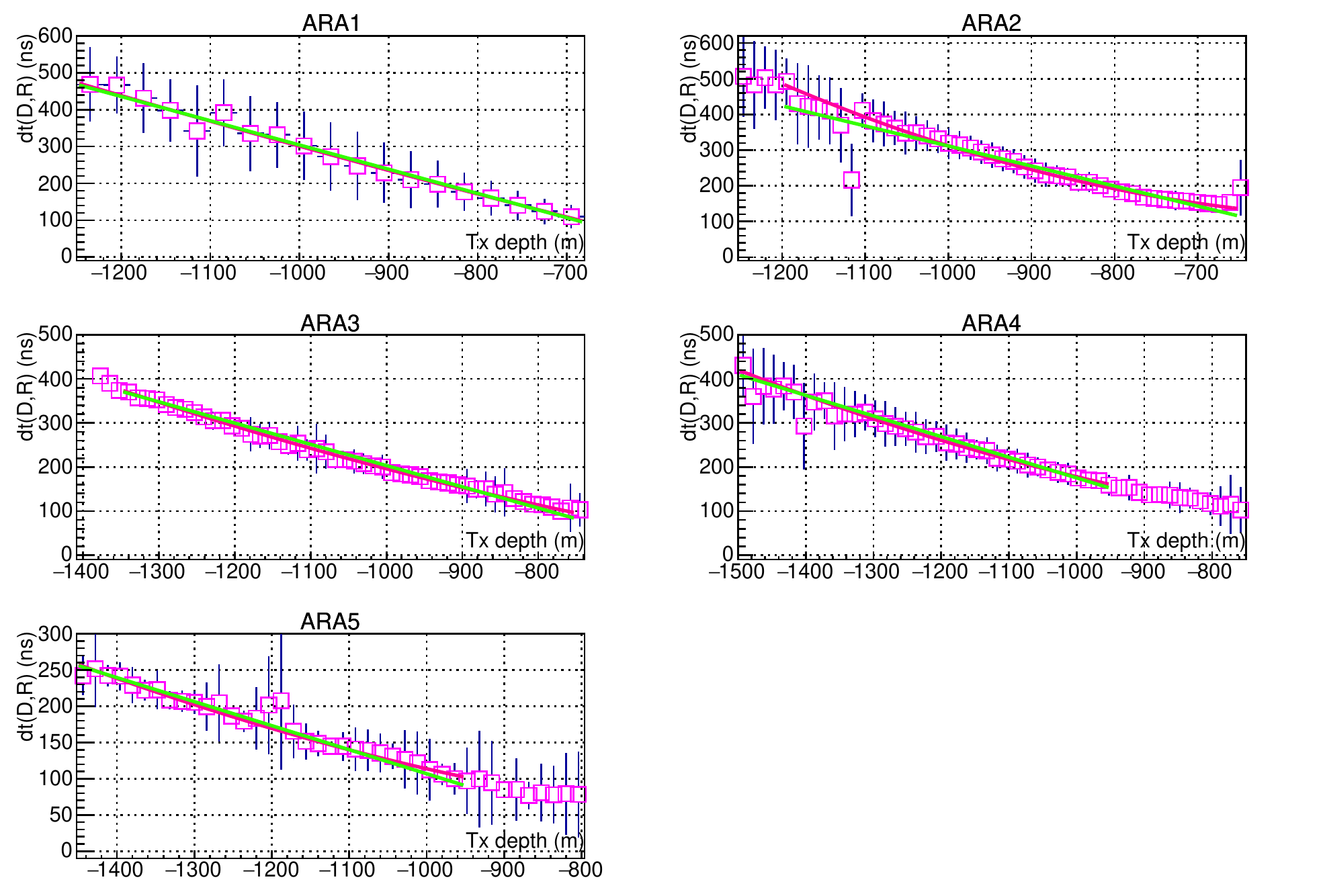}} 
\caption{Time differences (vertical axis, in units of ns) between Refracted (R) and Direct (D) pulses as a function of PVA depth into SPICE core (horizontal axis, in units of meters). Apparent 'flattening' of curves for small time difference values is an artifact of limitations of the double-pulse hit-finding algorithm.}
\label{fig:dtDR}
\end{figure}

\begin{figure}[htpb]
\includegraphics[width=0.65\textwidth]{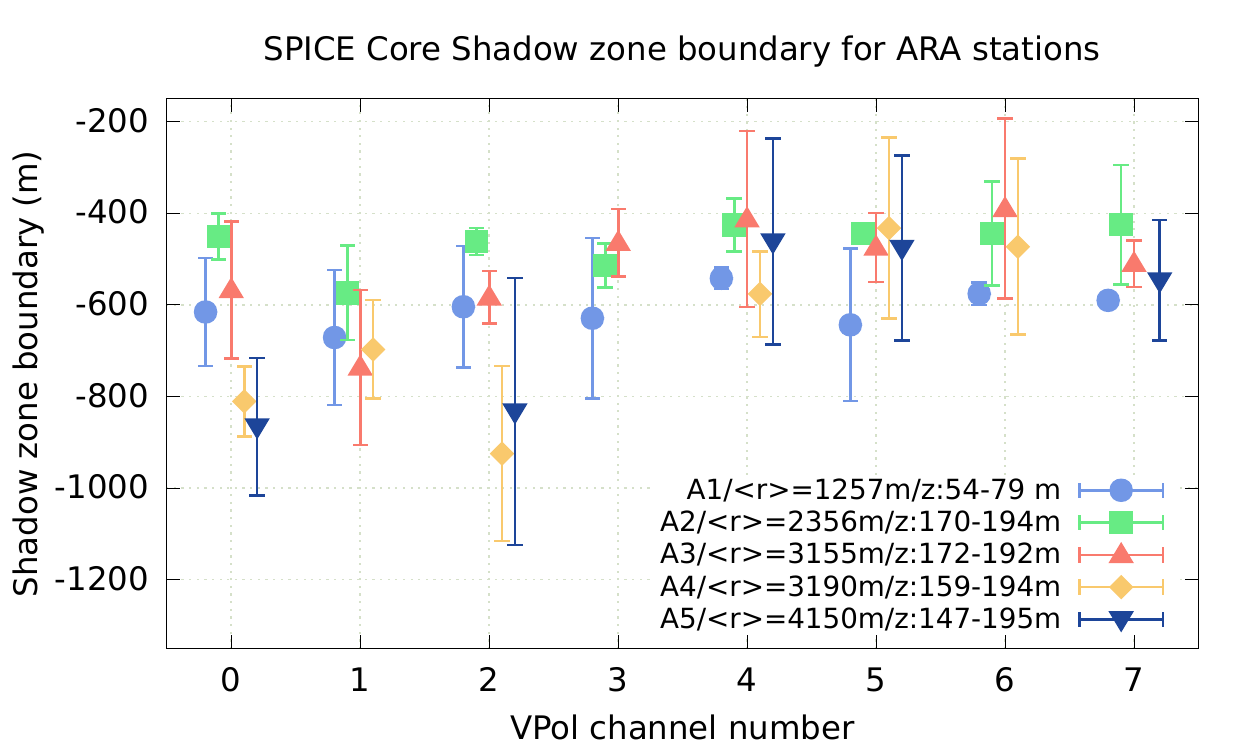}
\caption{Derived value of depth of shadow/visible zone boundary for VPol channels of ARA stations A1-A5, obtained by 
extrapolating $\delta_t(D,R)$ curves to zero time
difference. Average radial separation from station to SPICE core, and depth of VPol receivers (with channels 0-3 shallowest and channels 4-7 approximately 20-30 m deeper) are indicated in legend. Error bars indicate estimated uncertainties in shadow zone boundary extrapolation.}
\label{fig:shadow}
\end{figure}

For fixed receiver depth, the time difference between the Direct and Refracted pulses is expected to increase
roughly linearly with source depth once the transmitter emerges from the `shadowed' to the
`visible' region, qualitatively consistent with the trend observed in our data (Figures \ref{fig:PA} and \ref{fig:dtDR}).
Extrapolating the x-intercept of the (D,R) time differences allows us to infer the boundary between the shadowed vs. unshadowed regions, for a given 
receiver array.
The z-value corresponding to the depth of the `edge' of the shadow region for all of the channels for which we have
`good' ARA station data is presented in Figure \ref{fig:shadow}. Presented errors are estimated by considering the variation in x-intercept using
different parameterizations for the refractive index profile. 
As indicated in the Figures above, there is often measurable signal from the SPUNK PVA while the
PVA is still in the nominal `shadow' regime, corresponding to the upper few hundred meters of the ice sheet.
Measurable signals emanating from within the nominal shadow zone at both the South Pole and Moore's Bay have been previously reported elsewhere\cite{barwick2018observation} and may result in enhanced aperture for radio-based neutrino detectors.


\subsubsection{Implications for refractive index profile}
Note that, although the vertical separation between the four shallower VPol antennas relative to the four deeper
VPol antennas is only 20 meters, the (D,R) time difference separations for the two sets of antennas (shallower vs. deeper) are markedly different, suggesting that the $\delta_t(D,R)$ data might be used to extract the true {\tt n(z)} profile from a global fit. 
In principle, of course, if one has absolute timing relative to the signal emission time, the absolute transit time from transmitter to receiver, which depends on the integrated
travel time through a region of varying refractive index, might be used to extract
{\tt n(z)}. That measurement, however, is susceptible to combined systematic uncertainties in the geometry of the array laterally, as well as vertically.
The time differences between the measured D and R signal arrivals circumvent those uncertainties.
The constraints in our fits are the measured $\delta_t(D,R)$ time differences as well as the
extrapolated depth of the shadow/non-shadow transition. 
Quantitatively, the gradient in $\delta_t(D,R)[z_{Rx}]$ is of order 3--4 ns per meter in receiver depth, compared to the canonical 1 ns timing uncertainty in the ARA data. 

\begin{figure}[htpb]\centerline{\includegraphics[width=0.95\textwidth]{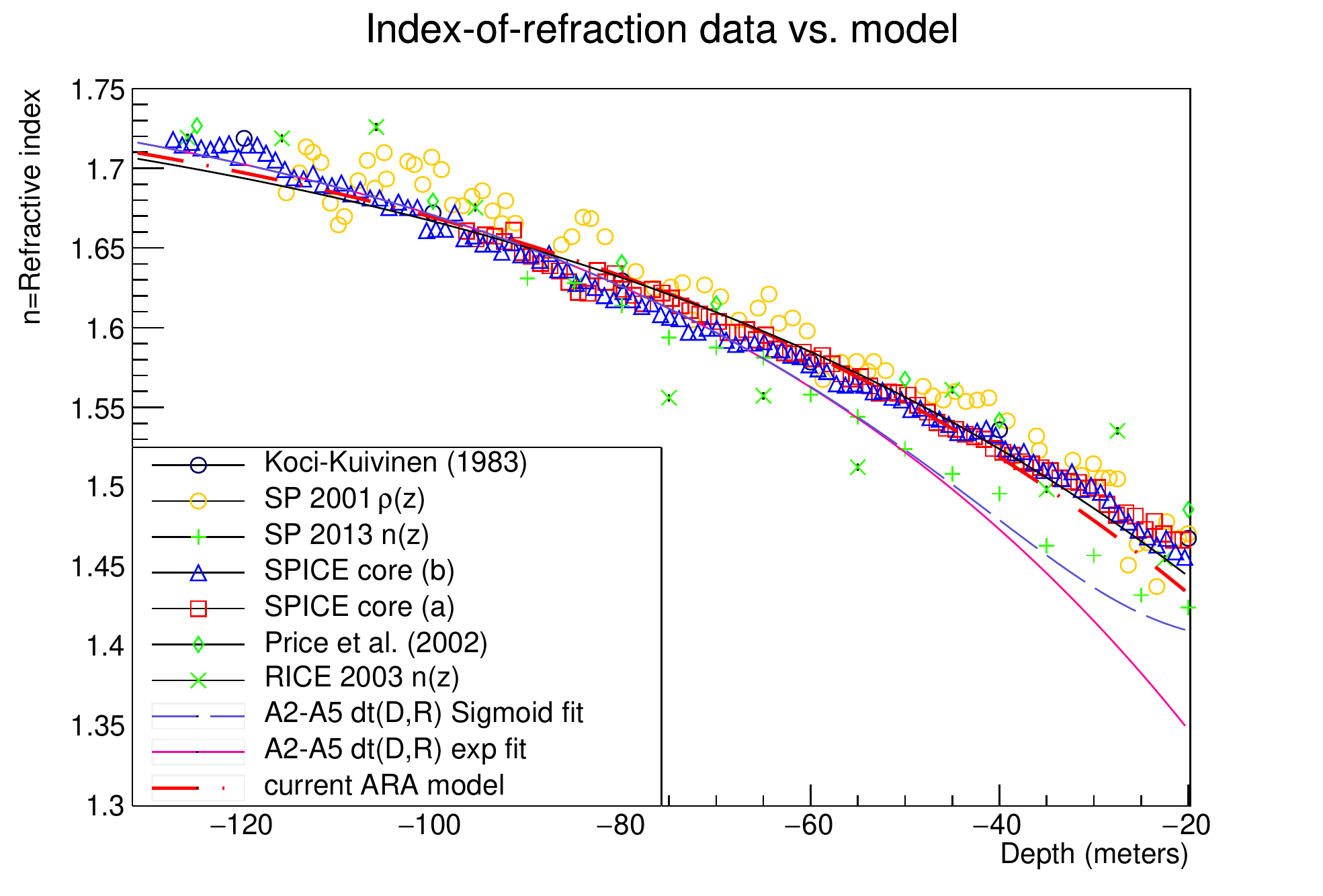}} 
\caption{Comparison of current ARA model (red, dashed), as well as sigmoid functional fit to ARA $\delta_t(D,R)$ data with existing South Polar density and refractive index data.
Density data are drawn from Koci-Kuivinen\cite{kuivinen1983237}, measurements based on a 2001 South Pole core\cite{van2008depth}, Price {\it et al.}\cite{price2002temperature}, and density measurements from the recent SPICE coring efforts (provided by Murat Aydin, private communication).}\label{fig:nvz}\end{figure}

The current preferred
ARA simulation (AraSim) model parameterizes the refractive index dependence on depth using the
exponential form {\tt n(z)}=1.78+Bexp(Cz), as expected for the density profile $\rho(z)$ of a fluid under its own
gravitational stress, with B=0.43 and C=0.0132 and depth negative [meters].
However, the existing South Pole density data suggest non-monotonic behavior in the upper 20 meters,
as also indicated by older RICE experimental measurements (e.g., Figure 3 of \cite{barwick2018observation}, and
also suggested by the observation of signals originating from `shadow'ed ice volumes at South Pole\cite{barwick2018observation}).
A sigmoid functional form allows a `flattening' of the refractive index
in the depth interval indicated by the RICE data, and asymptotes to the ARA functional form at
depths below z=-25 meters.
We initially determine the best-fit sigmoid functional
parameters using the $\delta_t(D,R)$ data from station A2. For A2, signals
refract and `turn over' at a maximum depth of 25 meters within the snow surface, for all SPICE geometries.
A comparison of the functional form extracted from
the fit to the A2 data with existing density data (extrapolated to {\tt n(z)} using the phenomenological ansatz
that the refractive index is a linear function of density {\tt n(z)}=1.+0.86$\rho(z)$), and also the current ARA model is
presented in Figure \ref{fig:nvz}. 

\begin{figure}[htpb]
\includegraphics[width=0.65\textwidth]{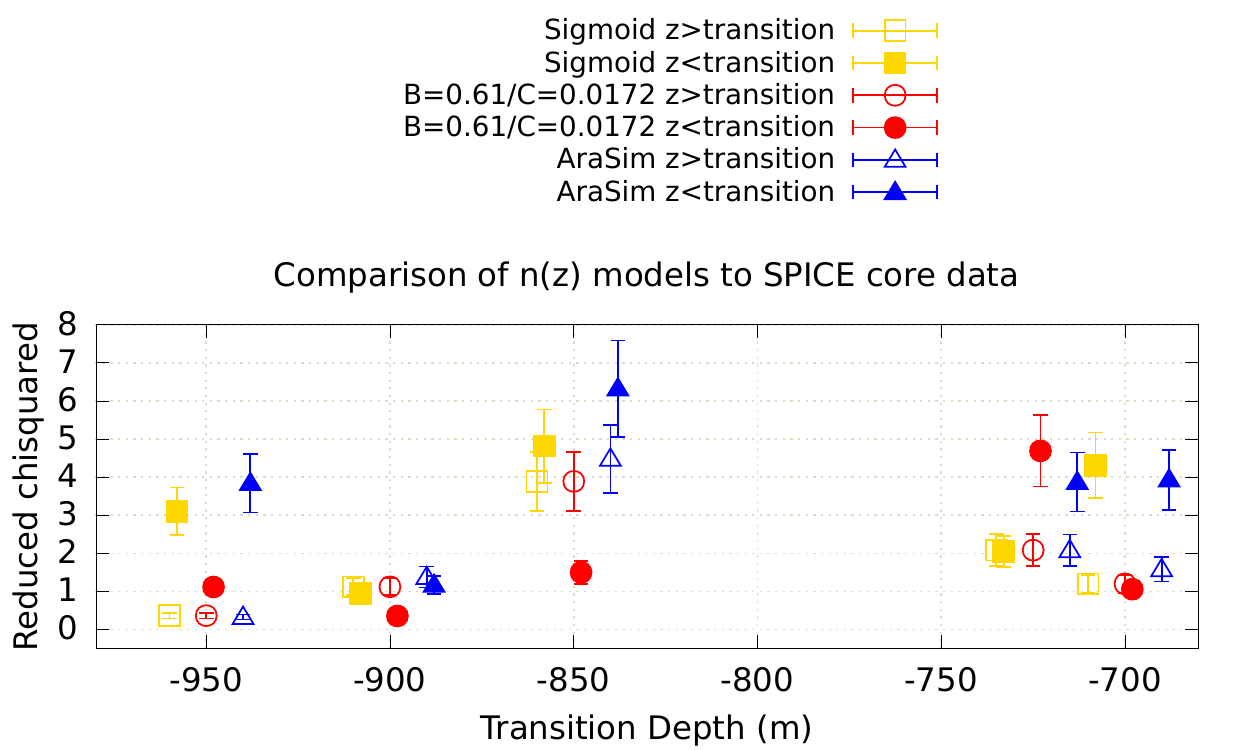}
\caption{Results of fitting three models for refractive index profile to SPICE core data. For these fits, we present results separately for the two indicated depth intervals.}
\label{fig:nz_fits}
\end{figure}

We have
compared the sigmoid vs. exponential-form
$\delta_t(D,R)$ predictions for double-pulses
observed using the deep pulsers IC1S and IC22S.  For the deeper stations A3--A5 as well
as the IC1S and IC22S pulsers,
our sigmoid model results in either equivalent, or improved chi-squared values 
relative to the Arasim exponential. However, for
station A1, for which rays refract to the receivers at shallow SPUNK depths,
or reflect for deeper source locations, Arasim is favored, suggesting the possibility
that the ice may not be as uniform over the multi-km scale as initially assumed. 
As an alternative, a modified exponential functional form using B=0.61 and C=0.0172 
gives superior results
compared to both the default AraSim {\tt n(z)} profile as well as a sigmoid for the majority of the data-sets used
for calibration, although it does not incorporate the measured 'break' in the South Polar refractive index at
shallow depths and underestimates the implied surface ice density. Figure \ref{fig:nz_fits} shows the goodness-of-fit comparison of the default ARA 
refractive index model (``AraSim'') compared to a revised exponential model, with B=0.61 and C=0.0172,
as well as the sigmoid model discussed previously. Our revised exponential model, overall, gives the best match to
observation, although it also results in an underestimate of the ice density at the surface.
In practice, the different parameterizations lead to very similar neutrino sensitivities.
As measured by the so-called `effective volume' (``$V_{eff}$'', or the equivalent volume of ice over which neutrino interactions are
measurable), the tested refractive index parameterizations give $V_{eff}$ estimates that agree to within 3\% of each other.

\section{Birefringence Measurements}
At radio frequencies, the bulk birefringence of polar ice arises due to the combined effect of the ice fabric (the orientation distribution of ice crystals) and the stretching of individual ice crystals. Whilst individual ice crystals have uniaxial symmetry, in the general case, polar ice is a biaxial medium (three different principal indices of refraction, which are aligned with the principal axes of the fabric orientation tensor). Given the two in-ice strains which break 3-space symmetry (vertical compression and horizontal ice flow), a reasonable ansatz aligns one principal axis in the direction of local ice flow (${\hat e}_1$), one axis vertical (${\hat e}_3$) and a third axis ${\hat e}_2$ perpendicular to both ${\hat e}_1$ and ${\hat e}_1$, as outlined previously. An electric field vector polarized along some arbitrary direction at the source is then projected into the three orthogonal planes defined by these three axes ($e_1e_2$, $e_1e_3$ and $e_2e_3$); the amplitude within each plane is uniquely determined by the condition that the Electric field be transverse to the Poynting vector (${\vec E}\times{\vec B}={\vec S}$). In this description of birefringence, each of the three allowed oscillation directions of the Electric-field components, separately confined to an orthogonal plane, then projects onto the axes defining that plane; those separately projected components subsequently propagate with the wavespeed specific to that polarization component. I.e., the E-field vector confined to the $e_1e_3$ plane is then resolved along the $e_1$ and $e_3$ axes; each of those polarization components propagate with velocity $c/n_1$ and $c/n_3$, respectively, with c=0.2998 m/ns. Note that transversality is not retained for these final, projected components, although the `parent' electric field polarization, confined to one of the allowed $e_ie_j$ planes, is perpendicular to both ${\vec B}$ and ${\vec S}$.
Formally, the propagation-vector ${\hat k}$, the Electric 
Displacement field ${\vec D}$ and the Magnetic Intensity ${\vec H}$ comprise a right-hand coordinate system:
${\hat D}\times{\hat H}={\hat k}$. Similarly, the direction of the Poynting Vector ${\vec S}$ is given by
${\hat E}\times{\hat H}={\hat S}$. For a birefringent medium, however, ${\hat D}$ is, in general, offset from ${\hat E}$ by some angle $\theta$, so that the signal propagation direction ${\vec k}$ is offset from the
energy flow direction ${\hat S}$ by the same angle $\theta$ (again violating transversality). The Poynting vector is therefore a `composite' of
energy transport along both the ordinary and extraordinary axes.

The RICE\cite{Besson:2010ww,Besson:2009zza}, ANITA\cite{besson2008situ}, and ARA\cite{allison2019measurement} 
experiments have all made extensive measurements of the
polarization dependence of radio-frequency signal amplitudes propagating through cold polar ice. The
first RICE 
measurements were limited to signals broadcast vertically into the ice from the surface and reflecting off internal
conducting layers (typically, ash resulting from volcanic eruptions)\cite{Besson:2010ww}. Waveforms were recorded as a function of the orientation of the polarization axis, taken at roughly 30-degree azimuthal increments in the horizontal plane, relative
to the local ice flow direction. Those first measurements indicated uniformity of echo return times for all measured polarizations,
to a depth of approximately 1.5 km into the ice sheet, at variance with volumetric scattering measurements made at Dome Fuji\cite{fujita2006radio}.
A second set of measurements were made for RF signals
propagating vertically through the entire 2.8 km thick ice sheet, and observed approximately 
30 microseconds later, after reflecting off the bedrock. Data from those studies 
demonstrated a maximal/minimal wavespeed for polarizations aligned parallel/perpendicular to the local
ice flow direction, with an asymmetry (i.e., time-difference divided by total transit time) of order 0.15\% between the two extrema\cite{Besson:2010ww}.
These results were somewhat at variance with simple expectations -- in the lower-half of the
ice sheet, the COF should be dominated by a uniaxial, vertical alignment, with no expected contribution to birefringence.
\message{We have conducted a re-analysis of the earlier RICE data, using the somewhat more sophisticated hit-finding and pattern recognition software developed for the ARA experiment. To a depth of approximately 1500 meters,vertically-propagating signals with orthogonal polarizations, in the horizontal plane, are confirmed to be simultaneous to within 1 ns for all azimuths.}

\begin{figure}[htpb]\centerline{\includegraphics[width=0.9\textwidth]{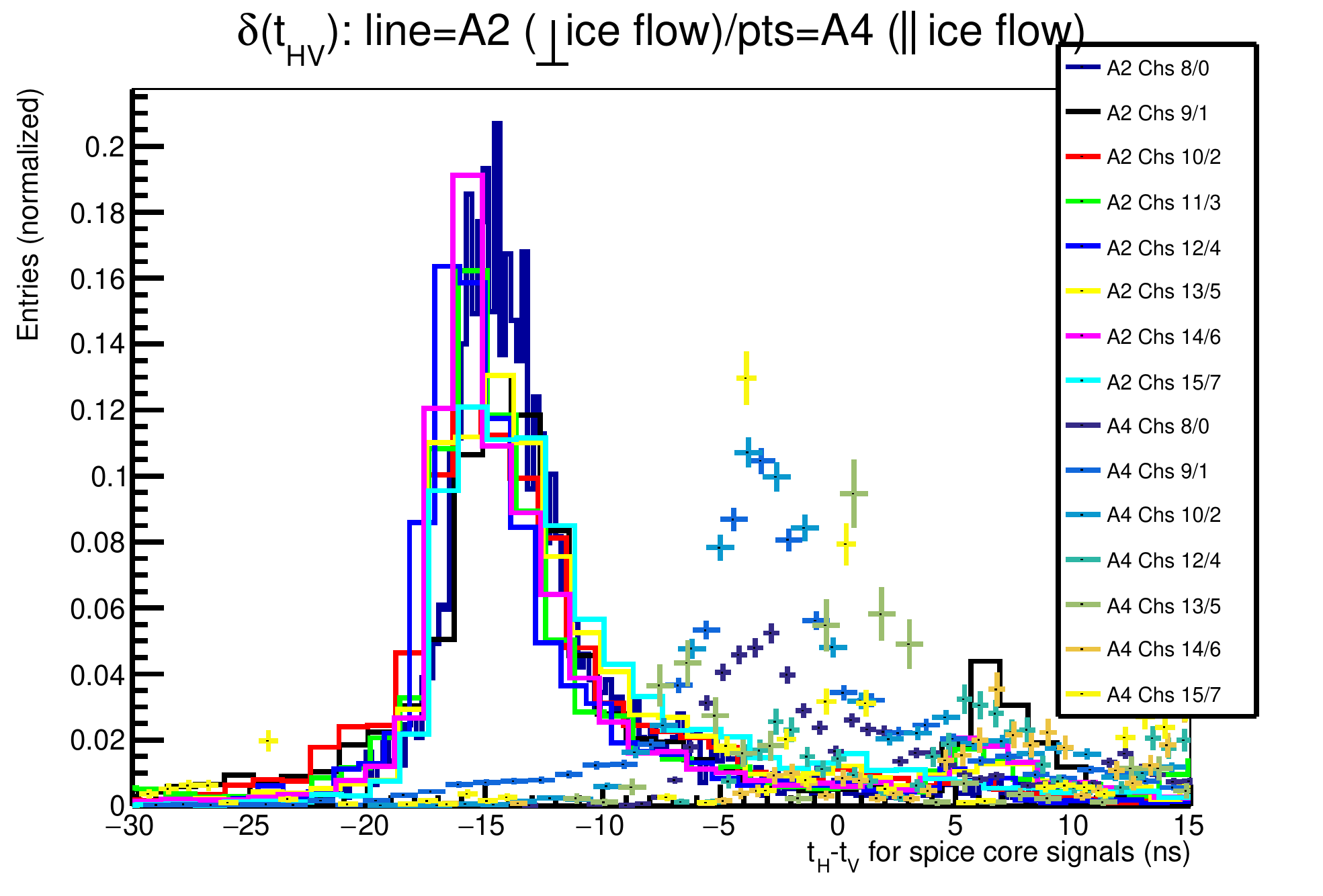}}\caption{Summary of birefringence measurements (transmitter=Spice Core pulser), comparing data from stations parallel vs. perpendicular to ice flow. We note that a) the resolution in the HPol-VPol time asymmetry 
is approximately 2-3 ns for most channels, and b) there are very few conspicuous 'outliers'.}\label{fig:birefsumsp}\end{figure}

\begin{figure}[htpb]\centerline{\includegraphics[width=0.9\textwidth]{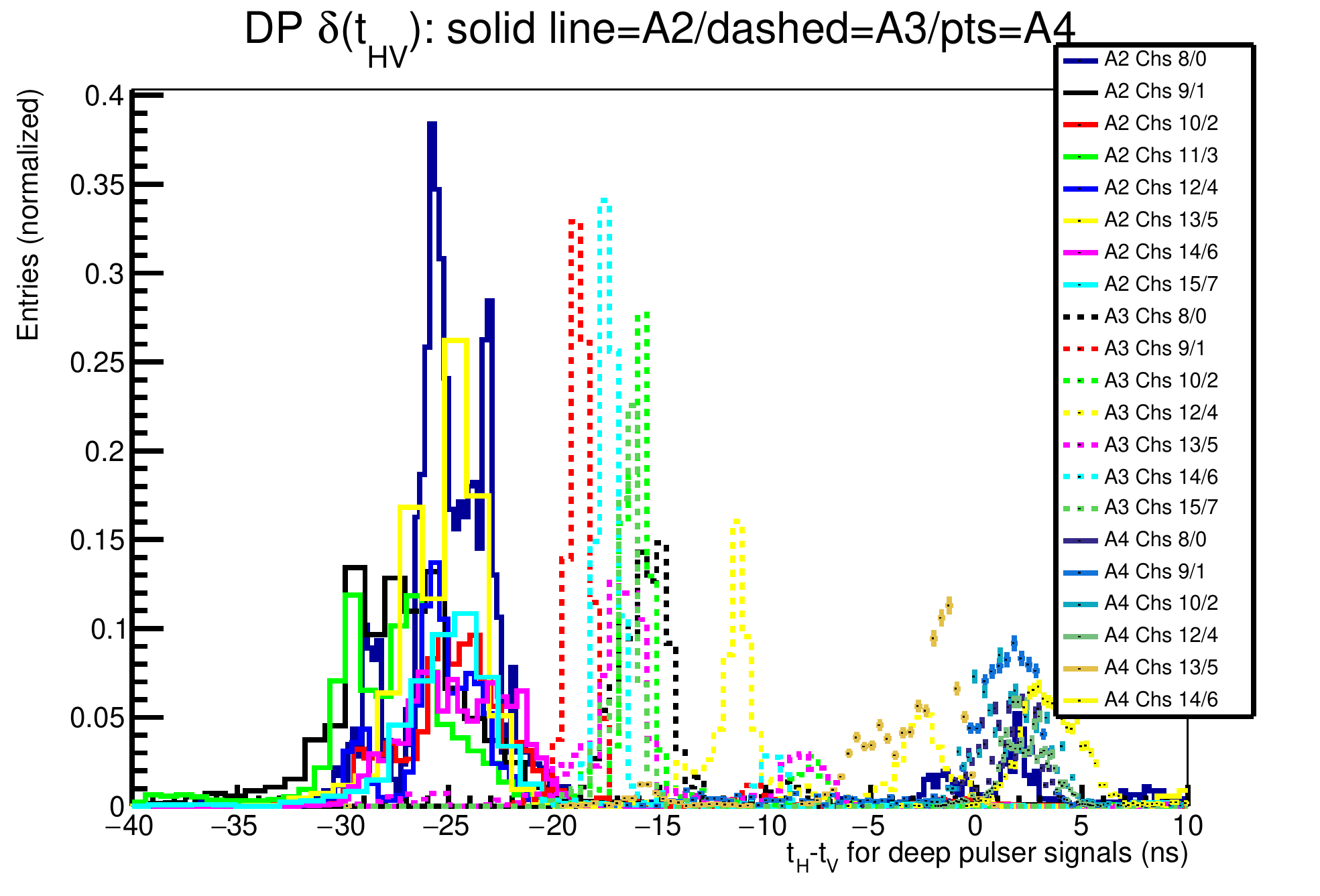}} 
\caption{Summary of birefringence measurements (transmitter=deep pulsers), comparing data from stations parallel vs. perpendicular to ice flow. We note that all stations are approximately equidistant from the source, and vary primarily in their orientation relative to the local ice flow direction.}\label{fig:birefsumdp}\end{figure}

\begin{figure}[htpb]\centerline{\includegraphics[width=0.9\textwidth]{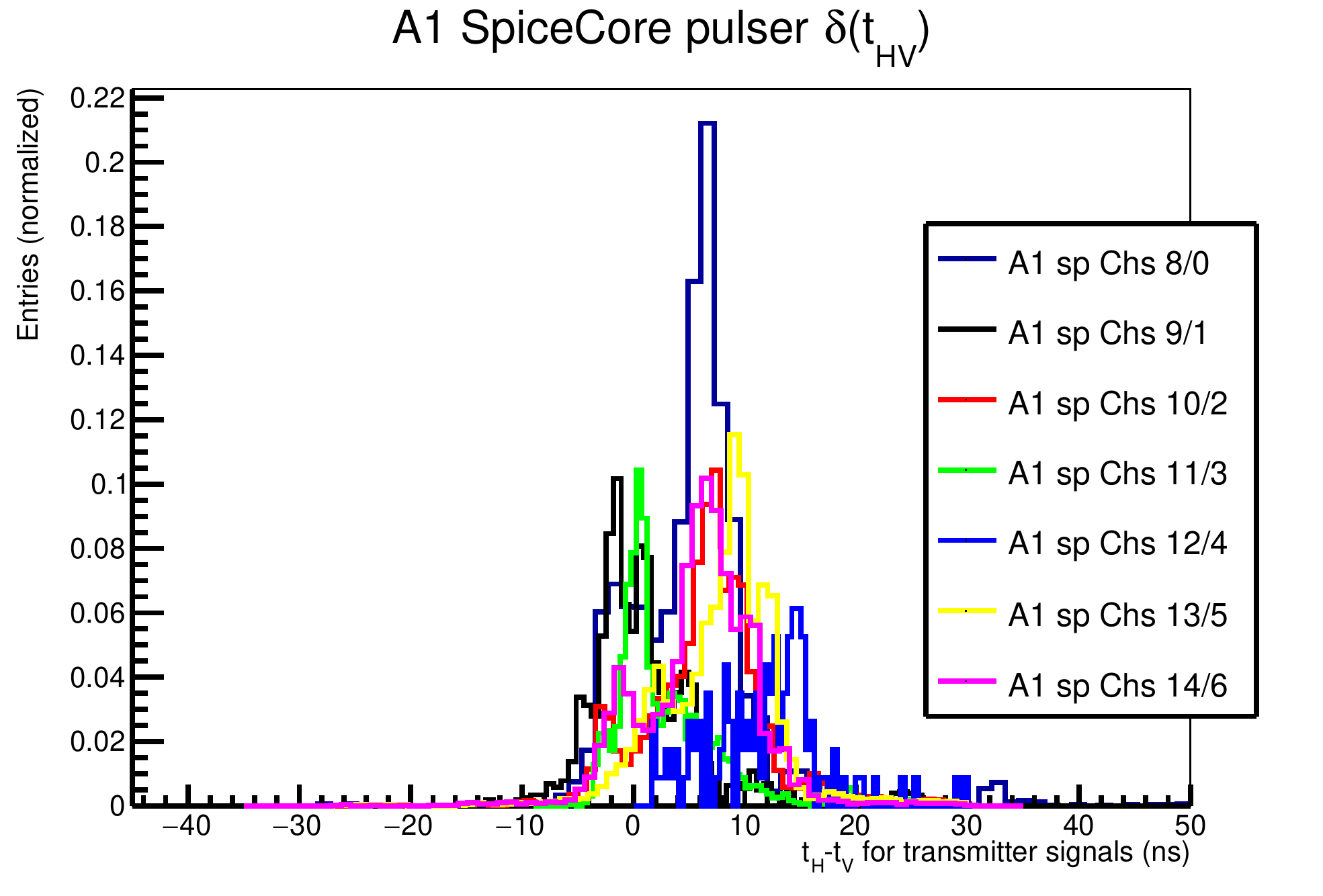}}\caption{Summary of birefringence measurements (transmitter=SPUNK pulser only) for ARA Station A1.}\label{fig:A1biref}\end{figure}

Our initial ARA time-difference asymmetry study for horizontally propagating signals\cite{allison2019measurement} used 2015
data collected using the IC1S and IC22S radio transmitters.
Since those data were collected, ARA station A4 was commissioned and analysis techniques improved. Broadcasts from
a transmitter in either the SPICE core hole, or the IceCube deep pulsers IC1S and IC22S, propagate
RF at directions roughly parallel relative to the local ice flow direction to A4, and therefore allow
a direct comparison of the asymmetry in the arrival time-differences between horizontally vs. vertically
polarized signals, for which receiver systematics (such as angle-of-incidence corrections) should largely cancel. Figures \ref{fig:birefsumsp}
and \ref{fig:birefsumdp} show the HPol-VPol time-difference asymmetries measured for broadcasts
from the SPICE core icehole and the deep pulsers, respectively. In both cases, although we observe
a time-difference asymmetry for station A2, we observe no obvious asymmetry for station A4 (propagation
parallel to local ice flow direction).
These conclusions are consistent with observations of SPICE signals propagating to ARA station A1 (Fig. \ref{fig:A1biref}), also aligned parallel to the local ice flow direction, albeit at approximately 40\% of the horizontal displacement relative to the source, compared to station A4. Interestingly, A1 data indicate an overall positive shift of HPol relative to VPol.
\message{Unfortunately, the A3 response to the SPUNK PVA was inadequate to yield reliable data.}
ARA Station A5, like Station A2, is approximately
perpendicular to the ice flow direction, however, owing to the large separation distance, the small signals observed on the HPol channels (for both SPUNK or Deep Pulser pulsing) disallow a reliable birefringence extraction.

\message{UPDATE WITH RE-ANALYSIS OF HRN2HRN PLUS ARIANNA}
We note that similar measurements for the ARA testbed, using the deep pulsers IC1S, IC22S, and IC1D, yield (H-V) preliminary time differences of
-9.1, -9.1 and -7.8 ns respectively, although systematic errors have not been assessed.
For those testbed measurements, the arrival times of the testbed VPol channels (3,6,2 and 4) were added and compared to the summed arrival times of the testbed HPol channels (0,1,5 and 7). Each of the four testbed strings has one VPol and one HPol channel, separated by 5 meters vertically, with HPol on top in two strings and VPol on top in the remaining two strings. A simple comparison of the sum of the VPol times vs. HPol times therefore directly gives the birefringent asymmetry, independent of arrival angle corrections. The testbed was de-commissioned in December, 2016, and was therefore not active during the
December, 2018 SPICE campaign.

To ensure that the observed offset from zero for A2 is not an artifact of a global mis-calibration of relative (H,V) channel delays, 
we have performed a similar relative-timing analysis using data collected from a surface pulser in both
2015 and 2018. For 2015 surface pulser broadcasts, we observe H/V timing offsets of
$-10.3\pm 3.6$ (A1), $-10.9\pm 5.4$ (A2), and $-13.1\pm 5.4$ ns (A3). For 2018 broadcasts,
the comparable values for A1, A2 and A4 are $-12.3\pm 4.1$, $-12.2\pm 8.6$ and $-5.6\pm 17.6$ ns, respectively.
We note that, for signals arriving at the critical elevation angle of $\approx 35^\circ$, the typical H/V vertical
pair separation of 2.5 meters translates to an expected time difference of approximately $-10.3$ ns, consistent with observations.

\subsection{Interpretation}
\begin{figure}[htpb]\centerline{\includegraphics[width=0.9\textwidth]{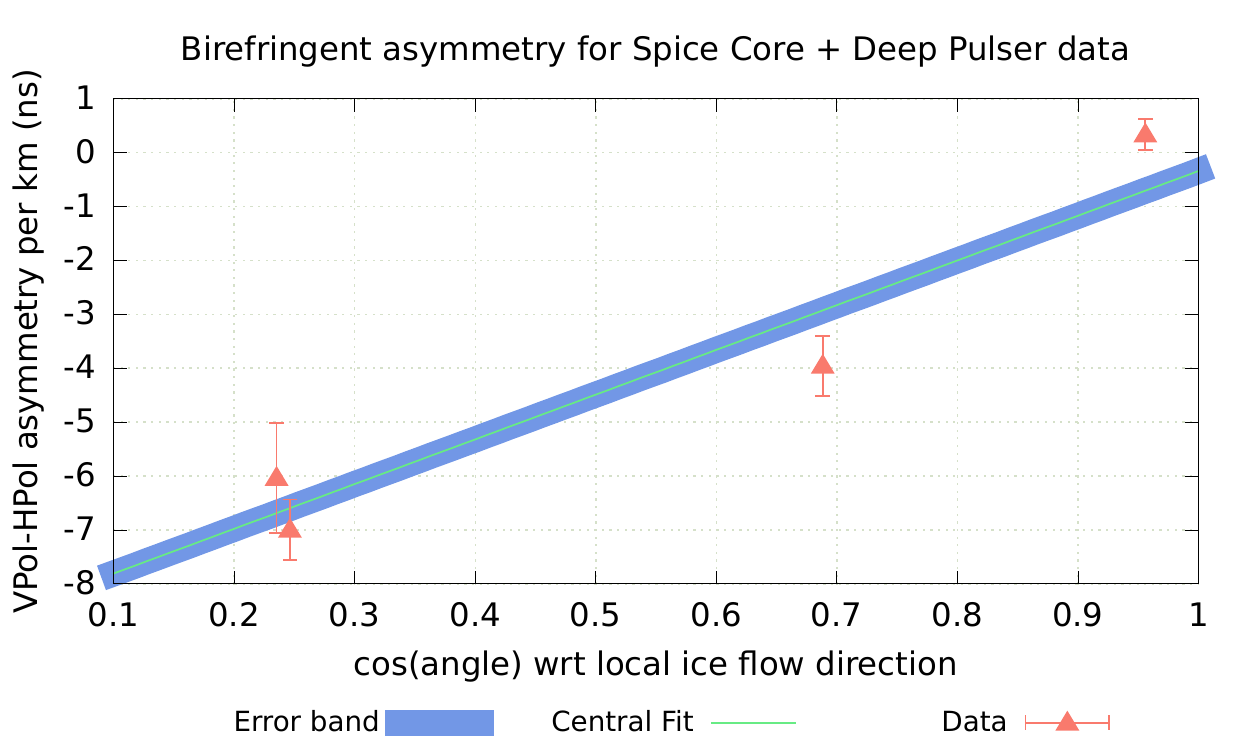}}\caption{Linear fit of measured birefringence asymmetry (per km), as function of signal propagation
angle relative to local ice flow direction (determined experimentally from interferometry). We fit our birefringence time-difference measurements to the functional form
$\delta_t(H-V) [{\rm ns/km}]=a\cos\theta+b$, with $\theta$ the
opening angle in the horizontal plane between the signal propagation direction and the local ice flow direction, and find
a=8.3$\pm$1.3, and b=-8.6$\pm$0.9 (errors combined statistical and systematic).}\label{fig:fitbiref}\end{figure}

We note the following: a) the ratio of birefringent asymmetries for A2 Deep Pulser pulsing vs. SPICE pulsing is approximately proportional to the pathlength difference for signals arriving from these two respective sources, b) although our A2 data was limited to the `visible' source range of 600-1100 meter depths, sub-dividing our data into $z<-$800 m vs. $z>-$800 m depth gives the same birefringent asymmetry, on average, to within one ns, consistent with a model where the ice core fabric does not vary significantly over this depth range.
The observed nearly linear variation of asymmetry with range suggests a linear parameterization of this dependence, which might then be used to 
translate a measured birefringent asymmetry from an unknown neutrino interaction point in-ice to estimate the range to interaction point (necessary for an estimate of the
neutrino energy). Such a linear fit is shown in Figure \ref{fig:fitbiref}.



\section{Radio-frequency attenuation length ($L_{atten}$) measurements}
Measurements of the absorptive component of the dielectric constant $\epsilon''$ require baselines comparable to several e-foldings of the signal amplitude to avoid being dominated by systematic uncertainties -- in the radio-frequency regime, the ice attenuation length is expected to increase by a factor of $\sim$5 over the temperature range 0$\to -$55 C, corresponding to the temperature variation between the bottom and the top of the ice sheet at the South Pole\cite{price2002temperature}, and consistent with the overall trend observed in Fig. \ref{fig:A5SNRvzTx}. Logistical considerations favor bi-static radar measurements, in which RF from a surface transmitter is reflected off the bedrock and then measured in a nearly co-located 
surface receiver (``bottom bounce''). The multi-kilometer, through-ice path, over a baseline $d_{ice}$ is then compared with a through-air path $d_{air}$. In that case, assuming 3-dimensional flux spreading and attributing all losses beyond the expected 1/r diminution of signal amplitude with distance to ice signal absorption, the antenna gain cancels in comparing the through-air with the through-ice signals and the attenuation length $L_{atten}$ can be extracted from the receiver signal voltages measured over a short baseline in air ($V_0$) vs. the signal voltage measured over a longer baseline in ice ($V_1$) via: 
$V_1/V_0=(d_{air}/d_{ice})exp(-(|d_{ice}-d_{air}|)/L_{atten})\times {\cal R}\times FFF$,  with
${\cal R}$ the unknown bedrock RF reflectivity (in the most conservative case, set to 1.0) and 
$FFF$ the Flux-Focusing-Factor, due to the variable refractive index profile with depth. For the typical bi-static
geometry, with nearly vertical propagation from/to the surface, 
the small-angle approximation gives $n_s\theta_s\sim n_i\theta_i$,
with $n_s$ the index-of-refraction of the antenna environment at the surface (presumably, roughly halfway
between air and surface snow/ice) and $n_i$ the asymptotic index-of-refraction below the firn; $n_i\approx 1.78$. In that limit, provided the firn layer is a relatively small fraction of the entire ice thickness, $FFF\approx n_i^2/n_s^2$.  \message{To determine the sensitivity to neutrino interactions, the attenuation length in the vicinity of the receiver antennas is most important.} In addition to systematic uncertainties due to flux-focusing and the bedrock reflectivity, these bottom bounce measurements measure the average attenuation over the entire thickness of the ice sheet, so the $L_{atten}$ dependence on depth must be corrected for. Previous estimates of the depth-dependent attenuation from bistatic measurements were therefore based on unfolding the depth-dependent attenuation length profile from the depth-averaged measurement, combining laboratory measurements of $L_{atten}(temperature)$ with the measured temperature vs. depth profile at South Pole\cite{price2002temperature}. \message{USE NEWEST DATA AS WELL?} The depth-averaged measurements are strongly dependent on the inferred ice temperature near the (relatively warm) bedrock, which produces the greatest ice absorption, and therefore dominates the unfolding.


Our broadcasts over long horizontal baselines permit an independent estimate of the 
radio-frequency attenuation in the upper (colder) half of the Antarctic ice sheet, corresponding to the bulk of the
proposed neutrino target volume for experiments based on detection of coherent Askaryan radiation. 
Both the SPICE core and the IceCube deep RF pulsers allow a more direct determination of $L_{atten}$ using
relative signal amplitudes observed at two stations ($V_1$ and $V_2$), and also knowing the pathlength
from source to receiver ($d_1$ and $d_2$, as shown in Table \ref{tab:geom}), via $V_1/V_2=(d_2/d_1)exp((d_2-d_1)/L_{atten})$; this
extraction is therefore insensitive to uncertainties in either the bedrock reflectivity or flux focusing. Broadcasts along the horizontal trajectories more typical of
neutrino signal geometries are also more sensitive to the possible `channeling' effects of
internal layers than bistatic vertical broadcasts, for which the internal
reflection/transmission coefficients are minimal/maximal.

\begin{figure}
\centerline{\includegraphics[width=0.9\textwidth]{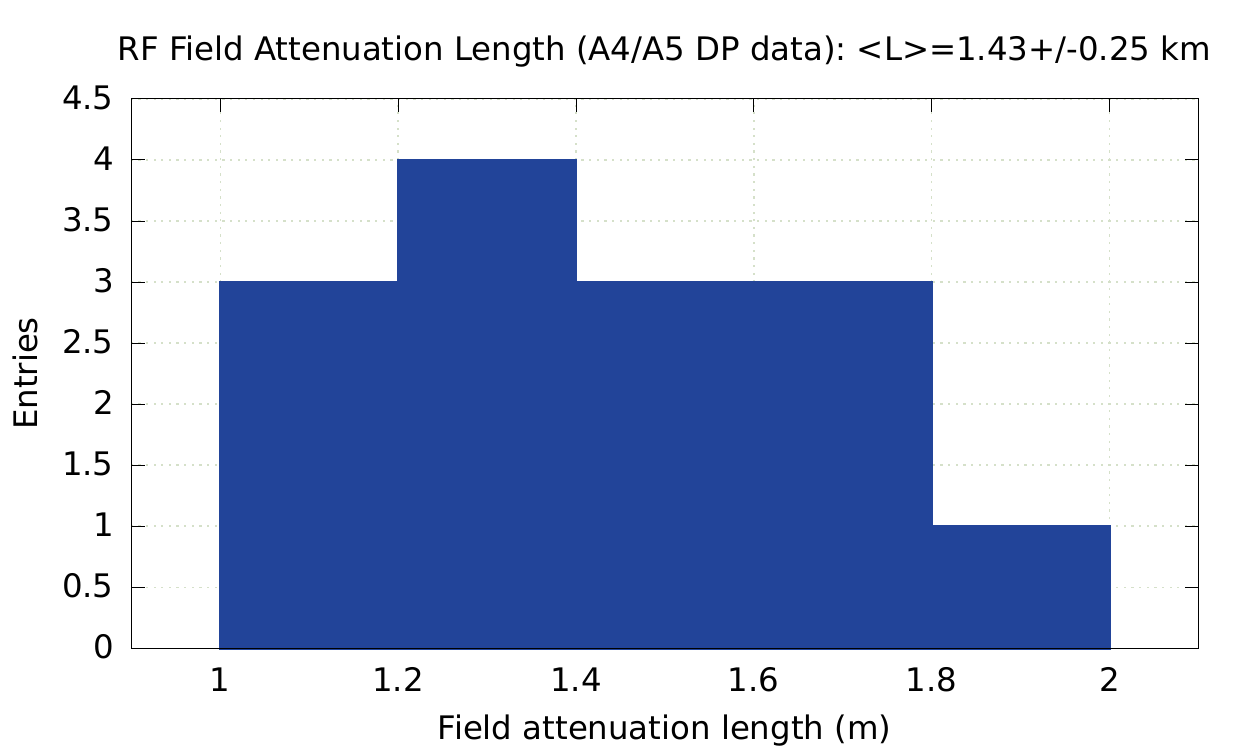}} 
\caption{Attenuation length based on relative VPol signal strengths observed in A5 relative to A4 using IceCube deep pulsers IC1S and IC22S as transmitters. To avoid
saturation of A4 waveforms, signal gain on A4 was attenuated by 12 dB (later corrected for in off-line analysis) for these runs.}
\label{fig:Latten45}
\end{figure}

Since the signal from the IC1S and IC22S transmitters is observed to saturate the A1, A2, and A3 receivers, measurements using those
deep pulsers
were restricted to ARA stations A4 and A5, which were equipped with attenuators at the input to the DAQ so that
reliable relative voltage values $V_1$ and $V_2$ could be extracted. The ARA Front End (ARAFE) modules used on 
stations A4 and A5 allow the user
to vary the signal attenuation at the ARA antenna output/DAQ input from zero to 12 dB, to mitigate saturation.
We make multiple estimates of the VPol attenuation
length, by comparing `same-antennas' A4 top-VPol channel 0 (TV0) to A5 TV0, etc. Although 
16 such combinations (8 VPol channels per receiver station $\times$ two deep pulsers; the weakness of HPol signals prohibits
an analogous HPol measurement) are, in principle,
possible, two combinations were excluded owing to non-functioning channels. The resulting distribution of calculated attenuation
lengths
is shown in Figure \ref{fig:Latten45}; this result is numerically consistent with $\sim$1.5 km previous estimates for $L_{atten}$ in the upper half of the
ice sheet\cite{allison2012design}.

For the non-saturating SPICE core data, we have taken a more general approach and produce an attenuation length distribution
for all possible `same-antenna' VPol pairs, for any two stations separated by at least two km. We note that this 
distribution is heavily weighted by station A5, which is furthest from the SPICE core. We additionally
separate our measurements by day, in order to probe possible differences in transmitter systematics.
We note that the data from Day 358, for which a piezo-based signal generator was employed, shows a 
somewhat longer attenuation length distribution than for the other three days. We regard those data as
most prone to systematic uncertainties than for the other days on which data were accumulated, owing to the
significantly less sharp signals, as well as the possibility of multiple after-pulses in a typical piezo-signal
waveform capture. Despite those uncertainties, the piezo data again indicate long attenuation lengths.
\begin{figure}
\centerline{\includegraphics[width=0.9\textwidth]{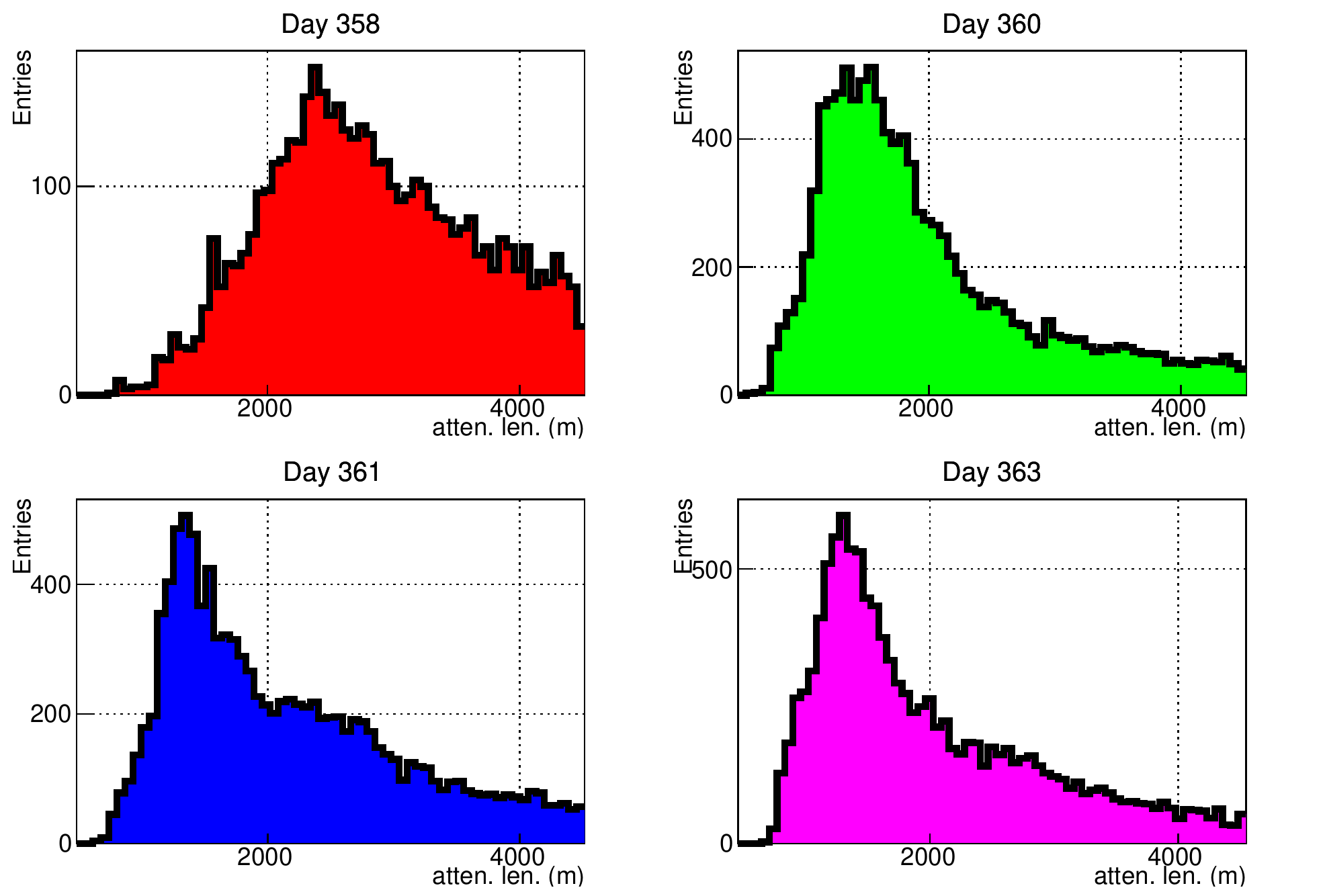}} 
\caption{Attenuation length distribution based on SPICE core pulsing for indicated days. Derived attenuation lengths are 
$L_{atten}=2377\pm 645$ (Day 358), 
$L_{atten}=1540\pm 361$ (Day 360), 
$L_{atten}=1348\pm 383$ (Day 361), and 
$L_{atten}=1302\pm 372$ (Day 363), respectively. Shown errors represent width of peak region of distribution.}
\label{fig:LattenSpice}
\end{figure}

If the amplitude resolution were infinite, each measurement would yield identical estimates for the
attenuation length. Although local calibration pulser events show clear fluctuations in the observed channel-to-channel signal amplitude,
Monte Carlo simulations indicate that the effect of finite resolution is to introduce some width to the
derived $L_{atten}$ distribution, but not to significantly
shift the peak value of that distribution. We 
correspondingly interpret the peak value of our $L_{atten}$ histograms as the radio-frequency
electric field attenuation length, averaged over the upper 200-1200 m of the South Polar ice sheet. 
Our measurements (Figure \ref{fig:LattenSpice}, corresponding to peak values of 
$L_{atten}$=2377, 1540, 1348 and 1302 meters, respectively) show general consistency with attenuation length estimates 
extracted from the deep pulsers, and are consistent with previous $>$1 km cold-ice 
radio-frequency $L_{atten}$
determinations\cite{barwick2005south,allison2012design}. 


\section{Systematics}
The calculations described herein are based on signal amplitude and signal arrival time 
measurements extracted from ARA receiver waveforms. Quantifying the variation in these parameters
is complicated by the intrinsic shot-to-shot variation in the SPUNK PVA transmitter output signal
strength and signal width, measured to be $\sim$40\%, and $\sim$35\%, respectively, 
in pre-deployment testing, with signal width/amplitude showing a general increase/decrease over time.
Loss of transmitter signal fidelity results in amplitude-dependent waveform slewing 
corresponding to approximately one
nanosecond of uncertainty in the hit-time definition; both birefringence and 
$\delta_t(D,R)$ estimates therefore have systematic uncertainties of order two ns, as evident from Fig.
\ref{fig:birefsumsp}.
This uncertainty is also typical of
saturated signals, since timing measurements are based on leading edges.
Attenuation length measurements are additionally sensitive to event-by-event variation in measured
signal strength for the ARA receiver stations. To isolate systematics at the
receiver end, the effect of transmitter-dependent variations
can be excluded by comparing the amplitudes and hit-times in two stations for the same
transmitter output pulse -- such events can be identified by requiring that the
global trigger time for two station triggers are coincident to within 100 microseconds,
corresponding to 5$\times$ the scale of the maximum signal transit time between the 
ARA stations registering triggers. From such coincidences (and corroborated by
the variation observed in local dedicated calibration pulsers), we estimate
the intrinsic variation in the measured individual-channel receiver signal amplitude to be of order 20\%. 
This value is smaller than the uncertainty in the channel-to-channel receiver gain, as evident from local calibration pulser signals.

The attenuation length distributions presented herein exhibit significant width. Monte Carlo 
simulations can be used to numerically interpret those widths. A simulation assuming infinite
precision in measured signal amplitude, as expected, yields a delta-function for the extracted
attenuation length distribution. We find that smearing our voltage amplitude distributions by a factor of 
2.4$\pm$0.4 (with the error shown indicative of the range of values required to match observations for the
four days shown), corresponding to an average voltage uncertainty of 3.8 dB, reproduces the width of the
attenuation length distributions shown. We emphasize here that that value includes not only the 
uncertainty in the intrinsic antenna plus data-acquisition electronics, but also uncertainties introduced
by propagation through ice from the SPUNK PVA to the receivers.

\section{Summary and Conclusions}
Data taken using either static, or movable radio-frequency transmitters at South Pole provide
an excellent opportunity for radioglaciology measurements and provide essential information to in-ice neutrino detectors.
Indeed, the SPICE core data sample described herein has very recently been used to test models of RF ice crystal geophysics\cite{jordan2019modeling},
quantify the sensitivity of the ARA detector to cosmogenic neutrinos\cite{allison2019constraints}, and quantify the polarization
response of the ARIANNA experiment\cite{anker2020probing}.
Herein, we present results based on multi-kilometer
propagation of radio-frequency signals broadcast through cold polar ice, with the goal of
quantifying the impact on neutrino detection experiments. We find:
\begin{itemize}
\item From a measurement of the difference in signal arrival time between Direct and Refracted rays, we have derived a range of
refractive index profiles which adequately fit the data. As shown elsewhere\cite{allison2019measurement}, this technique
allows a range-to-vertex estimate with uncertainty of order 10-15\%.
The error in the extracted refractive index profile implies a corresponding uncertainty in neutrino effective volume of order 2-3\%.
\item Multiple measurements of the signal arrival time as a function of polarization (`birefringence') demonstrate, for horizontally
propagating rays, a linear dependence of the birefringent asymmetry on range-to-vertex, once the signal incidence angle is determined by
standard interferometric reconstruction techniques. 
Although
ice birefringence presents additional challenges for data acquisition systems anticipated for next-generation
radio arrays, by inverting this linear dependence, the observed HPol/VPol time-difference asymmetry, combined with the additional geometric leverage afforded by observation
of both Direct and Refracted rays from an in-ice source also significantly improve the ability to estimate
the range to an in-ice neutrino vertex, which is itself essential for a neutrino energy estimate.
The range precision is geometry-dependent, and typically $\approx$15\%.
\item We demonstrate that cold polar ice has
the long radio-frequency attenuation length needed for cosmic ray neutrino sensitivity. 
These measurements, for the first time, are conducted over the long horizontal baselines characteristic of neutrino signals, unlike 
previous vertical `bounce' measurements.
\end{itemize}
With the cancellation of the 2020-21 South Polar field season, these are the last such measurements until such time as access to the SPICE core hole is restored; the timescale for those next measurements is uncertain.
\message{In the 2019-20 field season, additional measurements will be made with the SPUNK PVA, in an attempt to directly measure the frequency dependence of the polar ice attenuation length. That field season will also include dedicated measurements over the depth interval from +10 meters to -800 meters, to allow expanded studies, and tests of multiple-internal layer reflection models of shadow-zone propagation. In the summer of 2020, similar tests are anticipated for the Summit site in Greenland, for which the density and temperature profile are considerably different than South Pole.}

\acknowledgments
We thank the National Science Foundation for their generous support through Grant NSF OPP-1002483 and Grant NSF OPP-1359535. We further thank the Taiwan National Science Councils Vanguard Program: NSC 102-2628-M-002-010 and the Belgian F.R.S.-FNRS Grant4.4508.01. We are grateful to the U.S. National Science Foundation-Office of Polar Programs and the U.S. National Science Foundation-Physics Division. We also thank the University of Wisconsin Alumni Research Foundation, the University of Maryland and the Ohio State University for their support. Furthermore, we are grateful to the Raytheon Polar Services Corporation and the Antarctic Support Contractor, for field support. A. Connolly thanks the National Science Foundation for their support through CAREER award 1255557 and award 1806923, and also the Ohio Supercomputer Center. K. Hoffman likewise thanks the National Science Foundation for their support through CAREER award 0847658. A. Connolly, A. Karle, and J. Kelley thank the National Science Foundation for the support through BIGDATA Grant 1250720. B. A. Clark thanks the National Science Foundation for support through the Graduate Research Fellowship Program Award DGE-1343012. D. Besson and A. Novikov acknowledge the support from the MEPhI Academic Excellence Project (Contract No. 02.a03.21.0005) and the Megagrant 2013 program of Russia, via agreement 14.12.31.0006 from 24.06.2013. R. Nichol thanks the Leverhulme Trust for their support.
\section*{References}
\bibliography{/home/dbesson/updoc/Zref}
\bibliographystyle{unsrt} 

\end{document}